\documentclass[%
 preprint,
showpacs,
nofootinbib,
 amsmath,amssymb,
 prc,
 superscriptaddress
]{revtex4-1}

\usepackage{graphicx}
\usepackage{dcolumn}
\usepackage{bm}
\usepackage{braket}
\usepackage{caption}
\usepackage{subcaption}
\usepackage{mathrsfs}
\usepackage[utf8]{inputenc}
\usepackage{color}

\begin{document}


\title{
Alpha Clustering from the Quartet Model
}


\author{Dong Bai}
 \email{dbai@itp.ac.cn}
\affiliation{School of Physics, Nanjing University, Nanjing 210093, China}%

\author{Zhongzhou Ren}
\email{Corresponding Author: zren@tongji.edu.cn}
\affiliation{School of Physics Science and Engineering, Tongji University, Shanghai 200092, China}%
\affiliation{Key Laboratory of Advanced Micro-Structure Materials, Ministry of Education, Shanghai 200092, China}

\author{Gerd R\"opke}
\email{gerd.roepke@uni-rostock.de}
\affiliation{Institut f\"ur Physik, Universit\"at Rostock, D-18051 Rostock, Germany}
\affiliation{National Research Nuclear University (MEPhI), 115409 Moscow, Russia}


\date{\today}

\begin{abstract}

Alpha clustering in nuclei is considered with the quartet model (QM) where four valence nucleons (the quartet) move on the top of the core (daughter) nucleus. In the QM approach, it is assumed that the intrinsic wave function of the quartet is changed from the pure cluster configuration to the shell-model configuration when it crosses some critical radius and enters into the core nucleus. 
The QM approach could give not only the level scheme, the electromagnetic transition, the nuclear radius, but also the alpha-cluster formation probability. Numerical results are calculated for ${}^{20}$Ne, ${}^{44}$Ti, and ${}^{212}$Po, where a quartet moves on top of a double magic nucleus. Good agreement with experimental data and previous theoretical studies is obtained. The QM approach is a useful complement to the present phenomenological and microscopic models and could help deepen our understanding of alpha clustering across the nuclide chart.

\end{abstract}

\maketitle


\section{INTRODUCTION}

The study of alpha clustering could date back to Rutherfold's discovery of alpha decay, and nowadays it is still an important direction in modern nuclear physics. Previous studies show that alpha clustering could appear across the nuclide chart, from light and medium-mass elements to heavy and superheavy elements. Various phenomenological and microscopic models have been proposed in literature to describe various aspects of alpha clustering, see, e.g., Ref.~\cite{Beck:2010,Beck:2012,Beck:2014,Horiuchi:2012,Freer:2017gip,Delion:2010,Ren:2018xpt} for comprehensive reviews. Among them, the binary cluster model \cite{Buck:1990zz} and the quartetting wave function approach \cite{Ropke:2014wsa} are of special interest to the present work. 

The binary cluster model is a famous model originating from Gamow, Gurney and Condon's explanation of alpha decay in terms of quantum tunneling and marking the first application of quantum mechanics in the subatomic scale \cite{Gamow:1928,Gurney:1928}. Here, by ``binary cluster model'', we refer to a class of phenomenological models that regard the parent nucleus as a two-body system made of a tightly bound alpha particle and the core (daughter) nucleus. In literature, these models are also sometimes called as the local potential approach \cite{Michel:1998} or simply the cluster model \cite{Delion:2010}. With the alpha-core nuclear potential chosen properly to be, e.g., the $\text{WS}+\text{WS}^3$ potential \cite{Buck:1995zza} or the Woods-Saxon-Gaussian (WSG) potential \cite{Bai:2018hbe}, the binary cluster model is able to provide a systematic description of the level scheme, electromagnetic transition, and nuclear radius for alpha-cluster structures of various alpha$+$closed shell nuclei across the nuclide chart. In spite of these impressive phenomenological achievements, the binary cluster model by itself cannot give a meaningful estimation of the alpha-cluster formation probability \cite{Tonozuka:1979}, which is a key quantity to measure the strength of alpha clustering, as it has presumed that the parent state is composed solely of the alpha-cluster state from the very beginning. In other words, the alpha-cluster formation probability given by the binary cluster model should always be 100\%, which is not realistic. 

On the other hand, the quartetting wave function approach is proposed in 2014 as a new microscopic model to describe alpha clustering \cite{Ropke:2014wsa}, in which the parent nucleus is modeled by a compound system made of the core nucleus and four valence nucleons. The key feature of the quartetting wave function approach is that it allows a reliable estimation of the alpha-cluster formation probability with low computational costs. Unlike the binary cluster model, the four valence nucleons $\{ n_\uparrow, n_\downarrow, p_\uparrow, p_\downarrow \}$ in the quartetting wave function approach are assumed to form a tightly bound alpha particle only outside some critical radius determined by the Mott density and would merge into the shell-model states inside the critical radius. In the following, these four valence nucleons will be called as a quartet. The quartetting wave function approach has been adopted to study the ground-state alpha-cluster formation probabilities in various heavy and superheavy elements such as $^{212}$Po and its isotopes, as well as the light nucleus $^{20}$Ne, and the results agree well with previous microscopic calculations and empirical rules \cite{Xu:2015pvv,Xu:2017vyt,Ropke:2017att,Ropke:2017qck,Ropke:2018rkh}. These studies also inspired the very recent proposal of the cluster-daughter overlap as a new probe of ground-state alpha-cluster formation in medium-mass and heavy even-even nuclei in Ref.~\cite{Bai:2018}.

In this work, following the quartetting wave function approach, we would like to propose the quartet model (QM) as a new phenomenological model for alpha clustering and compare it with the binary cluster model. The QM approach attempts to provide a unified phenomenological description of various important properties of alpha clustering in alpha+closed shell nuclei across the nuclide chart, including not only the level scheme, the electromagnetic transition, the nuclear radius as discussed above for the binary cluster model, but also the alpha-cluster formation probability. 

The following parts of this paper are organized as follows. In Section \ref{QM}, we present the framework of the QM approach. In Section \ref{Examples}, as a proof of concept, the alpha-cluster structures of $^{20}$Ne, $^{44}$Ti, and $^{212}$Po are studied with the QM approach. Section \ref{Concl} ends this paper with conclusions and remarks.  

\section{QUARTET MODEL}
\label{QM}

This section presents the theoretical formalism for the alpha+closed shell nucleus, which is modeled by a quartet (four valence nucleons $\{ n_\uparrow, n_\downarrow,p_\uparrow, p_\downarrow\}$) moving on the top of the closed-shell core nucleus D in the QM approach. We assume the core D to be inert in our discussions, i.e., its state does not depend on the variables of the quartet. The parent wave function can be given by
\begin{align}
\Psi=\mathscr{A}\{\Psi^{(Q)}(\mathbf{r}_1,\cdots,\mathbf{r}_4)\Psi^{(D)}(\xi_D)\}.
\label{PaWF}
\end{align}
Here, $\mathbf{r}_i$ ($i=1,\cdots,4$) points from the center of mass (CM) of the core nucleus to the valence nucleon. $\xi_D$ corresponds to the intrinsic degree of freedom of the core nucleus. The quartet and core wave function $\Psi^{(Q)}(\mathbf{r}_i)$ and $\Psi^{(D)}(\xi_D)$ are assumed to be internally antisymmetrized respectively, and $\mathscr{A}$ is the interfragment antisymmetrization operator between the quartet and the core nucleus. In Eq.~\eqref{PaWF}, the wave functions $\Psi$ and $\Psi^{(D)}(\xi_D)$ are normalized.
Following Ref.~\cite{Ropke:2014wsa}, the quartet wave function $\Psi^{(Q)}(\mathbf{r}_i)$ is decomposed into the CM component and the intrinsic component
\begin{align}
\Psi^{(Q)}(\mathbf{r}_1,\cdots,\mathbf{r}_4)=\chi(\mathbf{R})\phi^{(Q)}(\mathbf{R},\xi_Q).
\label{CMDecomp}
\end{align}
Here, $\chi(\mathbf{R})$ is the CM component of the quartet wave function, and $\phi^{(Q)}(\mathbf{R},\xi_Q)$ is the intrinsic component of the quartet wave function, with $\mathbf{R}=(\mathbf{r}_1+\mathbf{r}_2+\mathbf{r}_3+\mathbf{r}_4)/4$ being the quartet CM coordinate, and $\xi_Q$ being the collection of three intrinsic coordinates of the quartet, which could be chosen to be, e.g., the Jacobi coordinates. The above decomposition is unique up to an arbitrary phase factor only, once the normalizations of $\Psi$, $\Psi^{(D)}$, and $\phi^{(Q)}$ are set down. With the help of Eq.~\eqref{CMDecomp}, the parent state is rewritten as
\begin{align}
\Psi=\int\mathrm{d}\mathbf{r}\,\chi(\mathbf{r})\Psi_\mathbf{r}(\xi_Q,\xi_D,\mathbf{R}),
\end{align}
where $\{\Psi_\mathbf{r}\}$ forms a basis labeled by the continuous subscript $\mathbf{r}$,
\begin{align}
\Psi_\mathbf{r}(\xi_Q,\xi_D,\mathbf{R})=\mathscr{A}\{\Psi^{(D)}(\xi_D)\phi^{(Q)}(\mathbf{R},\xi_Q)\delta(\mathbf{r}-\mathbf{R})\},
\end{align}
and $\chi(\mathbf{r})$ is the expansion coefficient. Based on these, the Schr\"odinger equation for 
$\varphi(\mathbf{r})\equiv\mathscr{N}^{1/2}\chi(\mathbf{r})$ is given by
\begin{align}
\mathscr{N}^{-1/2}\mathscr{H}\mathscr{N}^{-1/2}\varphi(\mathbf{r})=E\varphi(\mathbf{r}).
\label{NLSE}
\end{align}
$\mathscr{H}$ and $\mathscr{N}$ are integral operators obeying, e.g., $\mathscr{N}f(\mathbf{r})\equiv\int\mathrm{d}\mathbf{r}'N(\mathbf{r},\mathbf{r}')f(\mathbf{r}')$, and the corresponding kernels are given by $H(\mathbf{r},\mathbf{r}')=\braket{\Psi_\mathbf{r}|H|\Psi_{\mathbf{r}'}}$ and $N(\mathbf{r},\mathbf{r}')=\braket{\Psi_\mathbf{r}|\Psi_{\mathbf{r}'}}$. Compared with $\chi(\mathbf{r})$, the new wave function $\varphi(\mathbf{r})$ has the advantage to be normalized 
\begin{align}
(\varphi|\varphi)=(\chi|\mathscr{N}|\chi)=\braket{\Psi|\Psi}=1,
\end{align}
thus allowing the standard probability interpretation of quantum mechanics. Here, the angle brackets denote matrix elements with integrations over the physical coordinates $\xi_D$, $\xi_Q$, and $\mathbf{R}$, while the round brackets denote matrix elements with integrations over the parameter coordinate $\mathbf{r}$. It is easy to notice that the above formalism is a reminiscent of the resonating group method (RGM) \cite{Wheeler:1937zz} and the orthogonality condition model (OCM) \cite{Saito:1969zz}, with the intrinsic part $\phi^{(Q)}$ describing the more general quartet configuration rather than the pure alpha-cluster configuration. { 
The first application of the RGM for the alpha decay was given by Flie{\ss}bach \cite{Fliessbach:1975} who investigated antisymmetrization and normalization if the alpha cluster overlaps with the core nucleus.
} The intrinsic wave function of a quartet may change its form in dependence on the CM position $R$, from the alpha-like cluster state to an uncorrelated shell-model state. The Hamiltonian operator $\mathscr{N}^{-1/2}\mathscr{H}\mathscr{N}^{-1/2}$ is generally nonlocal. For practical calculations, it is convenient to approximate it by a local one,
\begin{align}
\mathscr{N}^{-1/2}\mathscr{H}\mathscr{N}^{-1/2}\sim H^\text{(QM)}\equiv-\frac{\hbar^2}{2\mu_\alpha}\nabla^2_\mathbf{r}+W(\mathbf{r}).
\end{align}
Here, $W(\mathbf{r})$ is the effective potential which could be determined phenomenologically by fitting, e.g., the observed level schemes of various alpha-cluster states, and $\mu_\alpha$ is the two-body reduced mass. The local-potential approximation is adopted widely in the phenomenological studies of nuclear cluster structures and heavy-ion collisions, and good agreements to the experimental data are achieved. As a result, Eq.~\eqref{NLSE} becomes
\begin{align}
-\frac{\hbar^2}{2\mu_\alpha}\nabla^2_\mathbf{r}\varphi(\mathbf{r})+W(\mathbf{r})\varphi(\mathbf{r})=E\varphi(\mathbf{r}).
\label{LSE}
\end{align}

The intrinsic wave function of the quartet $\phi^{(Q)}({\bf R},\xi_Q)$ is determined by the Schr\"odinger equation as shown in Ref.~\cite{Ropke:2014wsa}. 
In free space where the effects of the nuclear medium are absent, the solution is the well-known alpha cluster. In the high-density limit, the effective interaction between the constituents of the quartet becomes weak because of Pauli blocking owing to the surrounding nuclear medium. We don' t have a solution of the wave function for the intrinsic motion here, but make a phenomenological ansatz which is a superposition of both components, the alpha-cluster wave function and the product ansatz for the uncorrelated motion of the quartet nucleons, with coefficients depending on the CM position $\mathbf{R}$. 
Explicitly for the finite nuclei, the intrinsic quartet wave function $\phi^{(Q)}$ generally consists of the shell-model component $\phi_\text{SM}$ that considers the effects of the low-lying shell-model orbits and dominates in the small $|\mathbf{R}|$, and the cluster component $\phi_\text{Clus}$ that considers the effects of the high-lying shell-model orbits and dominates in the large $|\mathbf{R}|$. As known from the homogeneous nuclear matter,
the intrinsic wave function of the quartet changes abruptly its character at the Mott density $\rho_{\rm Mott} = 0.02917$ fm$^{-3}$ where the bound state merges with the continuum \cite{Ropke:2014wsa}. Inspired by this picture, to determine further the alpha-cluster formation probability, we make the simplification that
\begin{align}
\phi^{(Q)}(\mathbf{R},\xi_Q)=
\begin{cases}
\phi_\text{SM}(\mathbf{R},\xi_Q), & \ |\mathbf{R}|<R_\text{crit},\\
\\[-4ex]
\phi_\text{Clus}(\xi_Q), & \ |\mathbf{R}|>R_\text{crit}.
\end{cases}
\label{QWFA}
\end{align}
Here, $R_\text{crit}$ is the critical radius that separates approximately the shell-model-dominant region from the cluster-dominant region and is a free parameter to be determined later on. In other words, we assume that there is an abrupt change of the intrinsic structure of the quartet when it crosses the critical radius, i.e., the quartet is identified with the alpha particle only outside the critical radius and merges with the shell-model state inside the critical radius. This abrupt change 
 is a convenient approximation corresponding to the local density approximation frequently used in many-particle physics to describe inhomogeneous systems. For the later convenience, we also introduce the cluster basis $\{\widetilde{\Psi}_\mathbf{r}\}$ corresponding to the pure cluster configuration,
\begin{align}
\widetilde{\Psi}_\mathbf{r}(\xi_Q,\xi_D,\mathbf{R})=\mathscr{A}\{\Psi^{(D)}(\xi_D)\phi_\text{Clus}(\xi_Q)\delta(\mathbf{r}-\mathbf{R})\},
\end{align}
as well as the overlap integral operator $\widetilde{\mathscr{N}}$ and the corresponding kernel  $\widetilde{N}(\mathbf{r},\mathbf{r}')=\braket{\widetilde{\Psi}_\mathbf{r}|\widetilde{\Psi}_{\mathbf{r}'}}$. According to Eq.~\eqref{QWFA}, for $|\mathbf{r}|>R_\text{crit}$ and $|\mathbf{r}'|>R_\text{crit}$, we have
\begin{align}
\Psi_\mathbf{r}=\widetilde{\Psi}_\mathbf{r}, \quad\quad\quad \mathscr{N}=\widetilde{\mathscr{N}}, \quad\quad\quad N(\mathbf{r},\mathbf{r}')=\widetilde{N}(\mathbf{r},\mathbf{r}').
\label{Q2C}
\end{align}
{
The alpha-cluster formation probability $P_\alpha$ could then be obtained by \cite{Lovas:1998}
\begin{align}
P_\alpha&\equiv\int\mathrm{d}\mathbf{r}\braket{\Psi|\widetilde{\Psi}_\mathbf{r}}\widetilde{\mathscr{N}}^{-1}\braket{\widetilde{\Psi}_\mathbf{r}|\Psi}\label{DPA}\\
&=\int_{r<R_\text{crit}}\mathrm{d}\mathbf{r}\braket{\Psi|\widetilde{\Psi}_\mathbf{r}}\widetilde{\mathscr{N}}^{-1}\braket{\widetilde{\Psi}_\mathbf{r}|\Psi}+\int_{r>R_\text{crit}}\mathrm{d}\mathbf{r}\braket{\Psi|\widetilde{\Psi}_\mathbf{r}}\widetilde{\mathscr{N}}^{-1}\braket{\widetilde{\Psi}_\mathbf{r}|\Psi}\label{FPA}\\
&\approx \int_{r>R_\text{crit}}\mathrm{d}\mathbf{r}\braket{\Psi|\widetilde{\Psi}_\mathbf{r}}\widetilde{\mathscr{N}}^{-1}\braket{\widetilde{\Psi}_\mathbf{r}|\Psi}\label{FPA2}\\
&= \int_{r>R_\text{crit}}\mathrm{d}\mathbf{r}|\varphi(\mathbf{r})|^2.\label{SPA}
\end{align}
In Eq.~\eqref{DPA}, we project the parent state onto the alpha-clustering subspace. The operator $\widetilde{\mathscr{N}}^{-1}$ is needed here to take care of the nonorthonormality of the cluster basis $\{\widetilde{\Psi}_\mathbf{r}\}$. From Eq.~\eqref{FPA} to Eq.~\eqref{FPA2}, we have dropped out the first term that corresponds to the shell-model contribution to $P_\alpha$, as the alpha-cluster configuration is expected generally to be taken care of by the high-lying shell-model configurations rather than the low-lying ones spanned typically by single-particle orbits within the major shell only. This is also consistent with early studies on the shell-model approach to alpha decay, which show that it is typically smaller than the cluster contribution by about one order of magnitude. For instance, in Ref.~\cite{Varga:1992zz,Lovas:1998}, it is shown for the ground state of ${}^{212}\text{Po}={}^{208}\text{Pb}+\alpha$ that the alpha-cluster formation probability given by the low-lying shell-model components only (approximately given by the first term of Eq.~\eqref{FPA} in this work) is only about $3.7\times10^{-2}$, which is significantly smaller than the realistic alpha-cluster formation probability $P_\alpha=0.3$ (corresponding to Eq.~\eqref{DPA} in this work) as given by the cluster-configuration shell model. { It is reasonable to assume that similar relations hold also for ${}^{20}$Ne and ${}^{44}$Ti, { which are investigated in our work}. Also, the relation $\varphi(\mathbf{r})=\mathscr{N}^{-1/2}\braket{\Psi_\mathbf{r}|\Psi}$ and Eq.~\eqref{Q2C} have been used to obtain Eq.~\eqref{SPA}. As Eq.~\eqref{Q2C} holds only outside the critical radius under the assumption given by Eq.~\eqref{QWFA}, the first term in Eq.~\eqref{FPA} generally cannot be reduced to the integration of $|\varphi(r)|^2$ inside the critical radius.}
}

\section{EXAMPLES}
\label{Examples}

In Section \ref{QM}, we have worked out the theoretical formalism for the QM approach. To apply the QM approach in realistic studies, we have to make further decisions on how to choose the effective potential $W(\mathbf{r})$ and the critical radius $R_\text{crit}$. In this section, we would like to study various properties of alpha clustering in $^{20}\text{Ne}=\alpha+{}^{16}\text{O}$, $^{44}\text{Ti}=\alpha+{}^{40}\text{Ca}$, and $^{212}\text{Po}=\alpha+{}^{208}\text{Pb}$ to demonstrate the usefulness of the QM approach. For these targets, we choose the effective potential $W(\mathbf{r})$ to be the WSG nuclear potential in addition to the Coulomb potential and the centrifugal potential \cite{Bai:2018hbe}
\begin{align}
&W(r)=V_N(r)+V_C(r)+V_L(r),\label{WEP}\\
&V_N(r)=-\frac{V_0}{1+\exp[(r-R_D)/a]}\{1+\alpha\exp[-\beta(r-R_D)^2]\},\label{WSGP}\\
&V_C(r)=
\begin{cases}
\frac{Z_\alpha Z_c e^2}{r}, & \ r \ge R_D,\\
\\[-4ex]
\frac{Z_\alpha Z_c e^2}{2R_D}\left[3-\left(\frac{r}{R_D}\right)^2\right], & \ r<R_D.
\end{cases}
\\
&V_L(r)=\frac{\hbar^2}{2\mu_\alpha r^2}L(L+1).
\end{align}
Explicitly, we take the following parameters for the WSG potential in Eq.~\eqref{WSGP}:
\begin{align}
&V_0=203.3\text{ MeV},\quad a=0.73\text{ fm},\quad \alpha=-0.478, \quad \beta=0.054\text{ fm}^{-2},\nonumber\\
&R_D({}^{20}\text{Ne})=3.25\text{ fm},\quad R_D({}^{44}\text{Ti})=4.61\text{ fm},\quad R_D({}^{212}\text{Po})=6.73\text{ fm}.
\end{align}
Compared with the original Woods-Saxon potential, which is designed to describe the mean field on a nucleon moving in the nucleus, the WSG potential introduces an additional Gaussian term which describes the modification if the nucleons form an alpha particle \cite{Bai:2018hbe}. 
We consider it here as a phenomenological part in analogy to the optical potential to achieve a better agreement with experimental data. The microscopic origin of the WSG potential is quite complicated and is an important question to be answered in future works. As discussed in Ref.~\cite{Ropke:2014wsa}, it might be related to the complexity of the nuclear forces of the nucleons, as well as the Pauli blocking felt by the quartet when it penetrates the core nucleus, which is generally a nonlocal effect exchanging the nucleons in the core nucleus and those in the quartet. { Noticeably, the use of an additional Gaussian correction to the mean field is also proposed in Ref.~\cite{Delion:2013tla}, which aims to describe the alpha decay at the microscopic level.} In the local density approximation, such an additional term is obtained within the quartetting wave function approach \cite{Ropke:2014wsa} and is solely depending on the local density $\rho_D( R)$. The WSG potential is also featured by its universality, i.e., it could provide satisfactory descriptions for the alpha-cluster structures in $^{20}$Ne, $^{44}$Ti, and $^{212}$Po using almost the same parameter set with only the radius parameter $R_D$ being modified correspondingly. It is also important to have a physical understanding of this property. 

\begin{figure}
\centering
\includegraphics[width=0.6\textwidth]{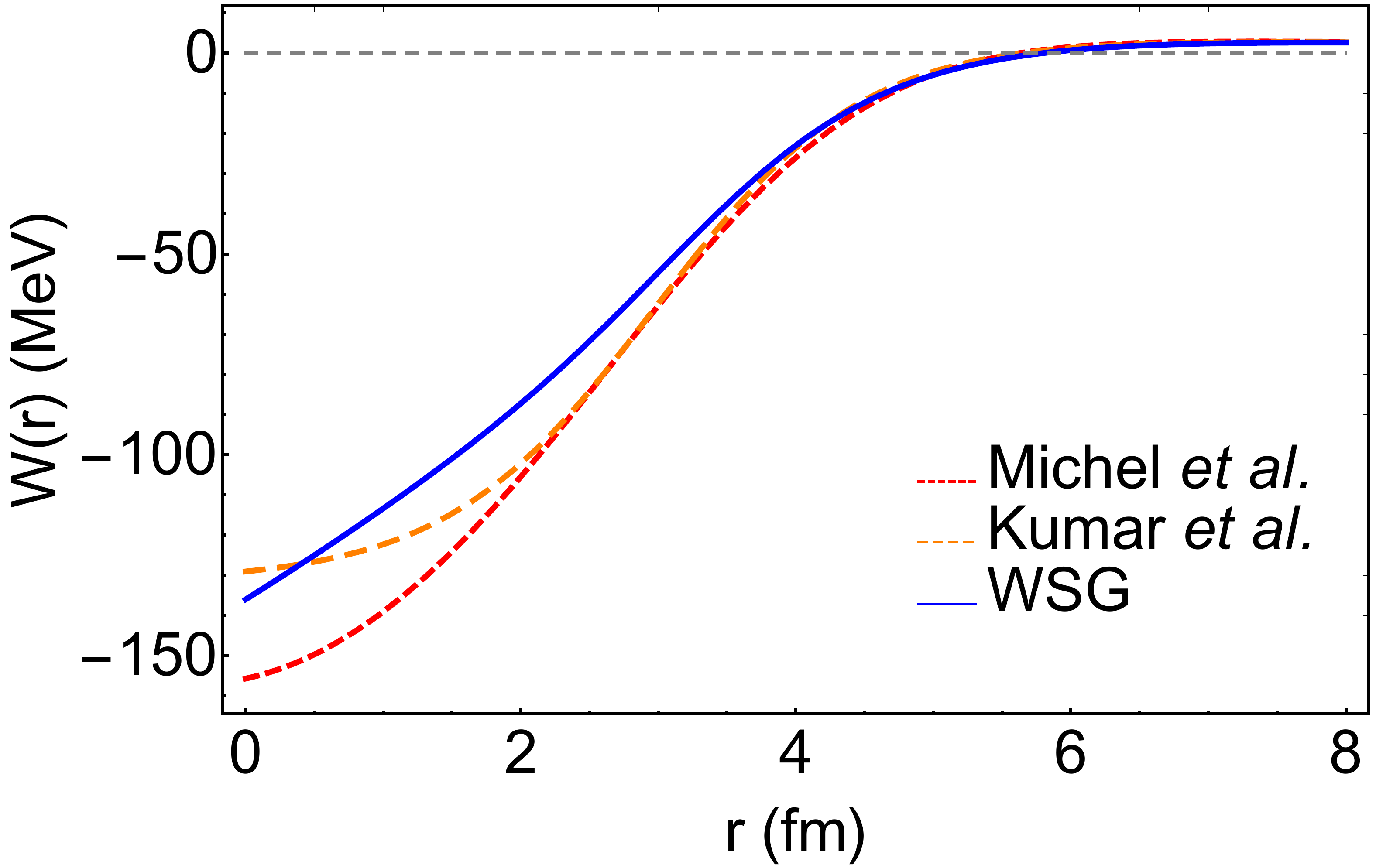}
\caption{The comparison of the WSG potential between the quartet and $^{16}$O with the real parts of various optical potentials including the Michel potential \cite{Michel:1983zz}, and the Kumar potential \cite{Kumar:2006ntk}.}
\label{WSG}
\end{figure}

In Fig.~\ref{WSG}, we compare the WSG potential with the real parts of various optical potentials between the alpha particle and ${}^{16}$O obtained by analyzing the nuclear reaction data. It is found that in the surface region ($r>3.5$ fm), the WSG potential matches well with the Michel potential \cite{Michel:1983zz} and the Kumar potential \cite{Kumar:2006ntk}, which provides extra supports for the validity of the WSG potential.

A second input is the critical radius $R_\text{crit}$, which is considered here as an empirical parameter. It is varied around the benchmark value determined by matching the tail of the core-nucleus density profile with the Mott density for the homogeneous nuclear matter $\rho_\text{Mott}=0.02917\text{ fm}^{-3}$. 
To determine the benchmark value of $R_\text{crit}$, we take the density profiles for the doubly magic nuclei $^{16}$O, $^{40}$Ca, and $^{208}$Pb from Ref.~\cite{DeJager:1987qc}, which are also summarized in Appendix \ref{DenPro}. In Fig.~\ref{DensityProfile}, we plot the density profiles for these core nuclei, where the benchmark values of the critical radii determined by the Mott density of the homogeneous matter are denoted by $R_\text{Mott}$. Explicitly, we have $R_\text{Mott}=3.34$ fm for ${}^{16}$O, $R_\text{Mott}=4.50$ fm for ${}^{40}$Ca, and $R_\text{Mott}=7.74$ fm for ${}^{208}$Pb. 

To solve the Schr\"odinger equation for the CM motion of the quartet $\varphi(r)$, using the effective potential $W(r)$ given by Eq.~\eqref{WEP}, we have to obey the Pauli principle inside the core nucleus.
In principle, we can introduce an effective Pauli repulsive potential as in the quartetting wave function approach, which, when combined with the nuclear potential, could give a nice potential pocket at the nuclear surface. Here, we adopt another method by using the Wildermuth condition \cite{Wildermuth:1977} to impose additional requirements on the node number in the physical wave function. For $^{20}$Ne, it is required that the physical quartet wave functions satisfy $G\equiv2n+L=8$, with $n$ being the number of nodes in the radial wave function and $L$ being the orbital angular momentum. For $^{44}$Ti and $^{212}$Po, we have $G=12$ and $G=18$, respectively. Such an ansatz has been adopted by various studies and is able to give reliable results \cite{Buck:1995zza,Bai:2018hbe}.

%

\begin{figure}

\centering

\begin{subfigure}[b]{\textwidth}
\centering
\includegraphics[width=0.45\textwidth]{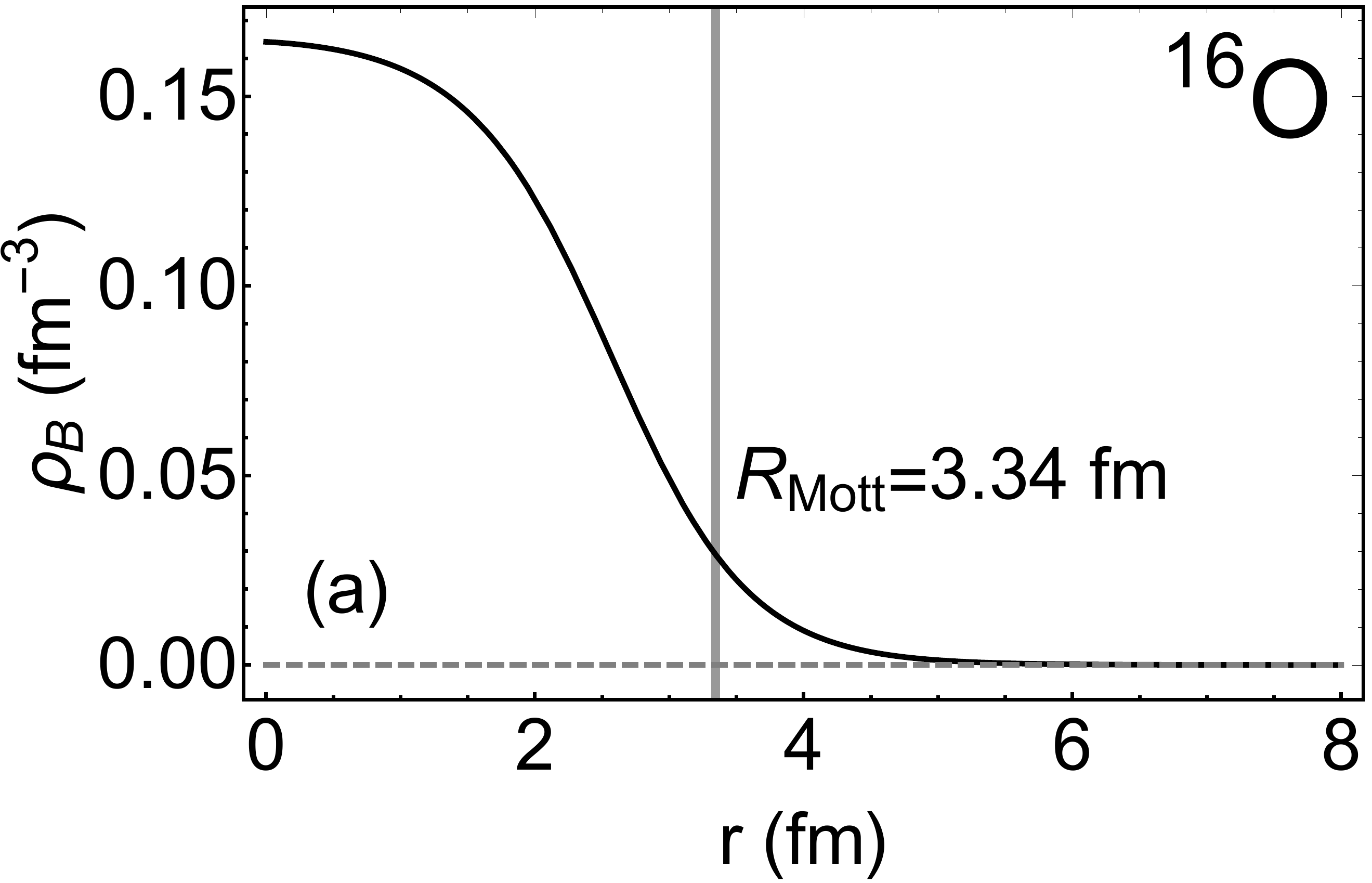}
\end{subfigure}

\begin{subfigure}[b]{\textwidth}
\centering
\includegraphics[width=0.45\textwidth]{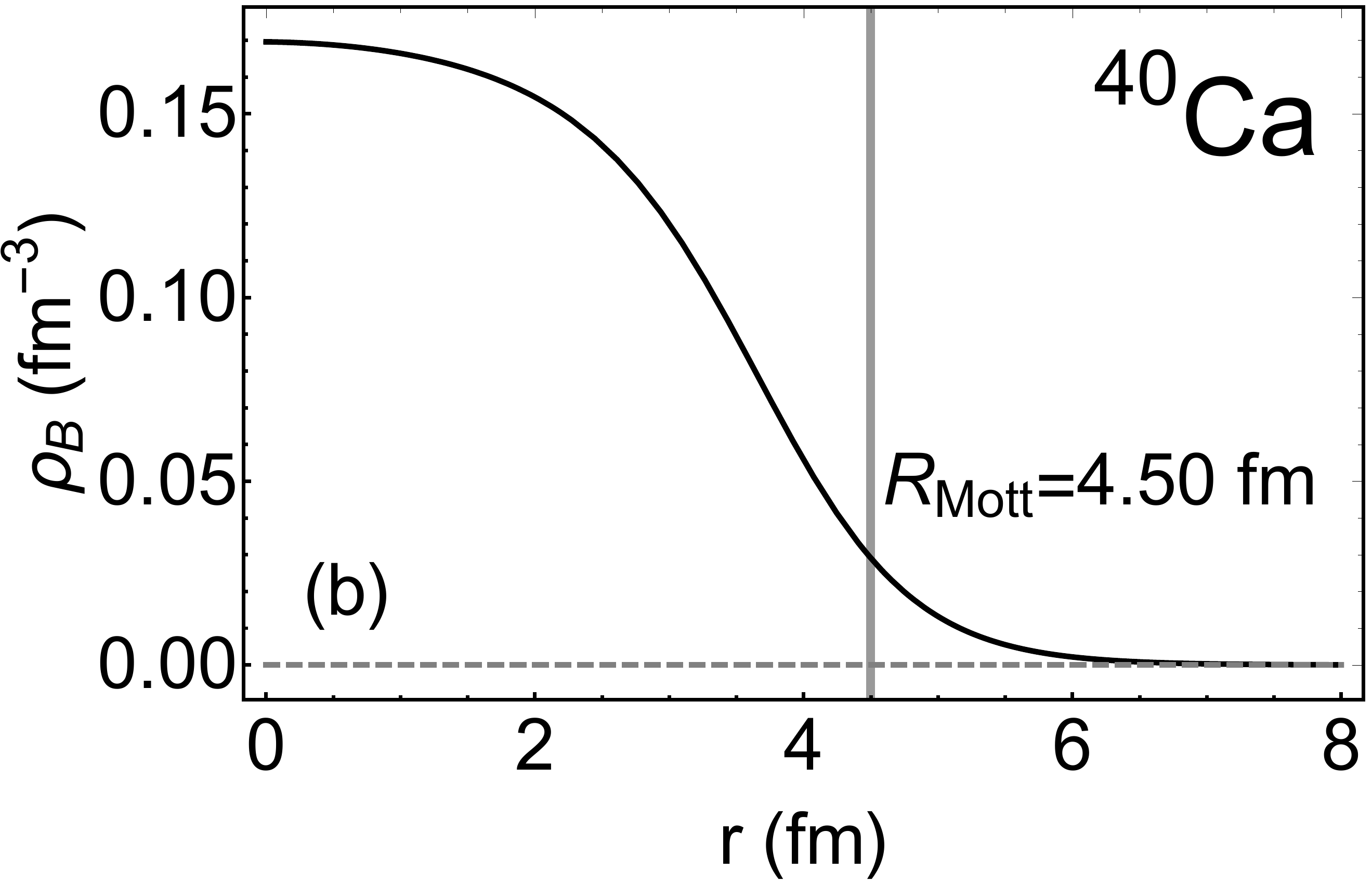}
\end{subfigure}

\begin{subfigure}[b]{\textwidth}
\centering
\includegraphics[width=0.45\textwidth]{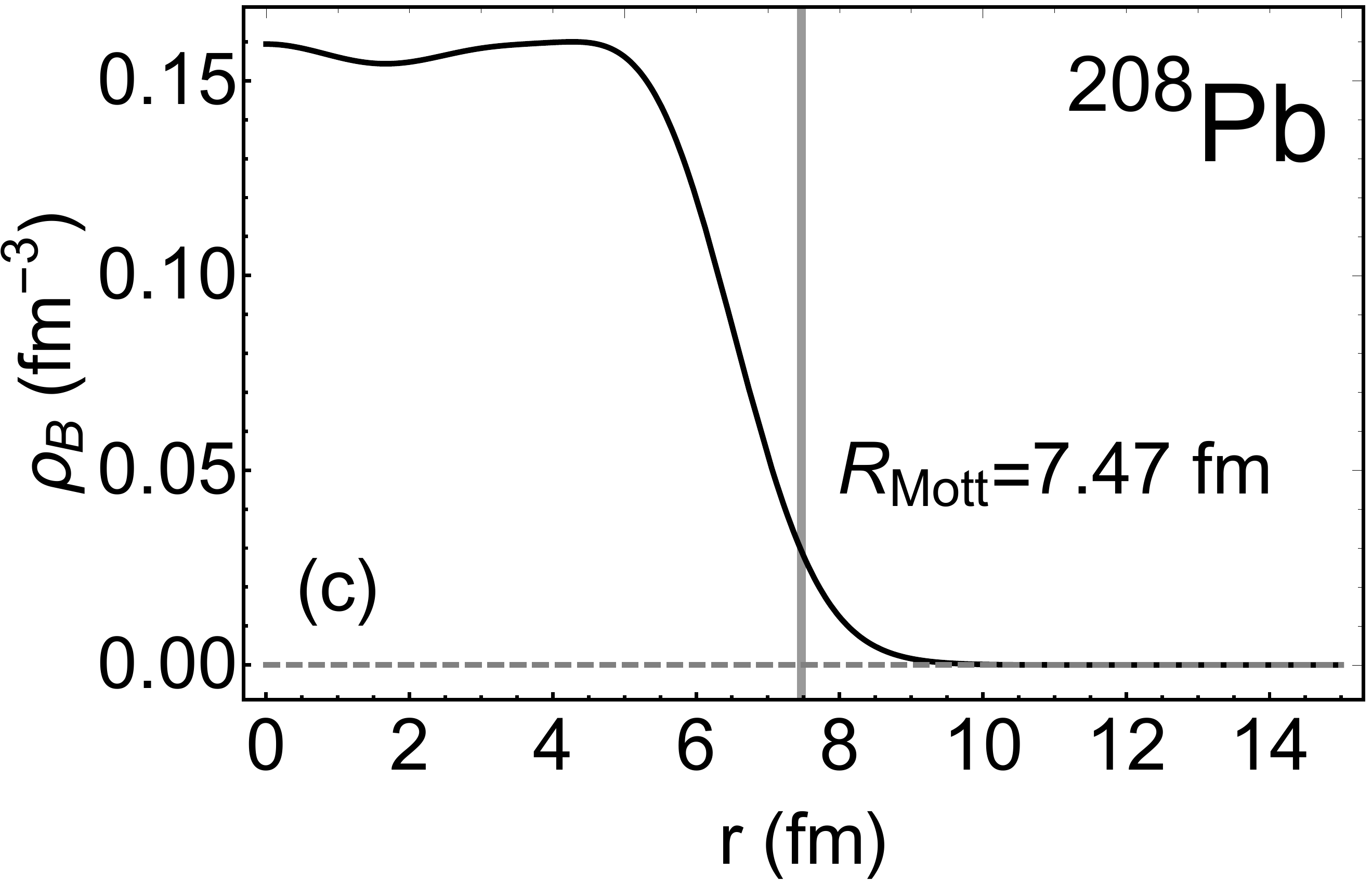}
\end{subfigure}

\caption{Plots of the density profiles for the doubly magic nuclei $^{16}$O, $^{40}$Ca, and $^{208}$Pb taken from Ref.~\cite{DeJager:1987qc} and summarized in Appendix \ref{DenPro}. $R_\text{Mott}$ denotes the benchmark value of the critical radius determined by matching the tails of the density profiles with the Mott density for the homogeneous nuclear matter $\rho_\text{Mott}=0.02917\text{ fm}^{-3}$, which is about one fifth of the nuclear saturation density.}
\label{DensityProfile}
\end{figure}

\begin{figure}

\centering

\begin{subfigure}[b]{\textwidth}
\centering
\includegraphics[width=0.45\textwidth]{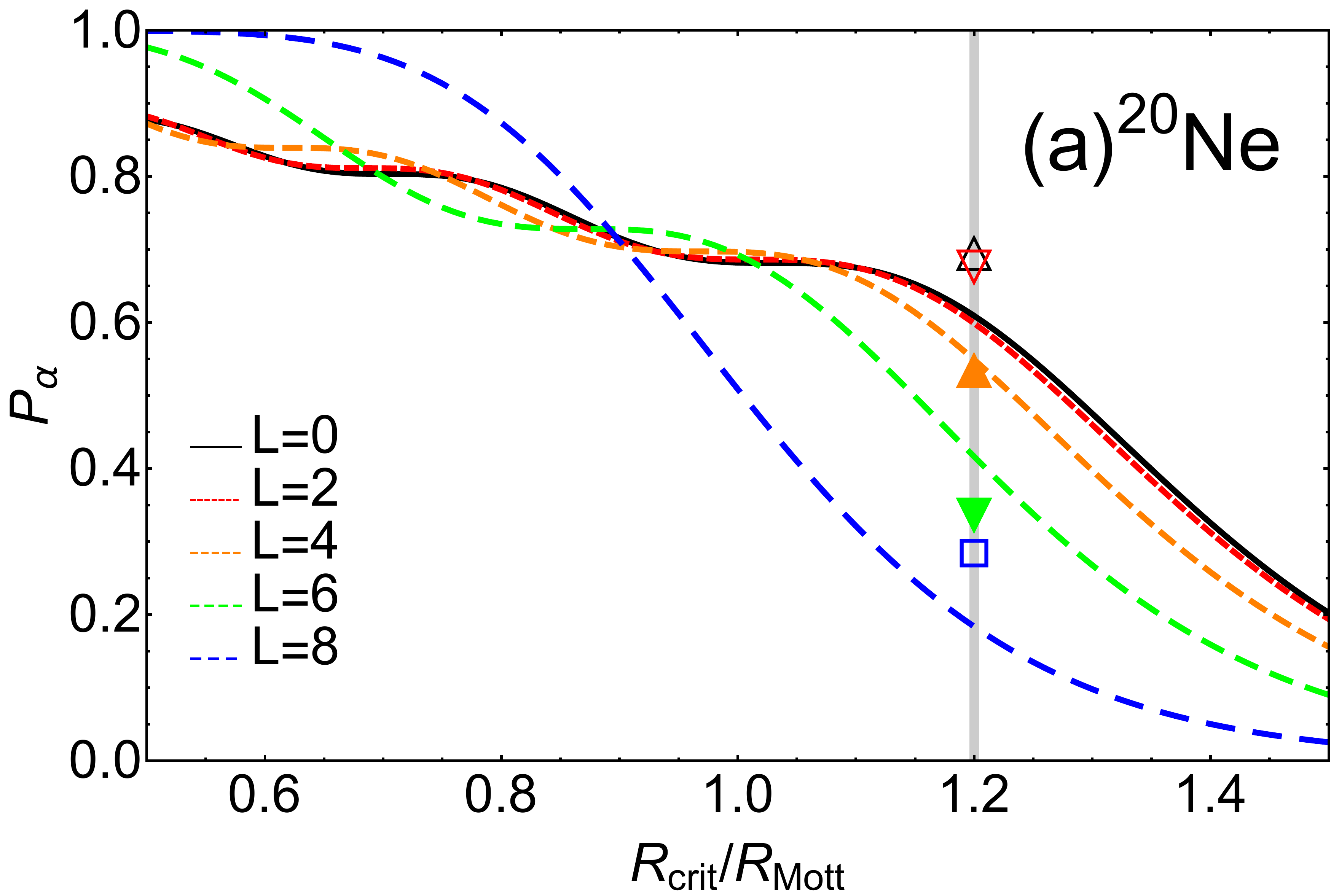}
\includegraphics[width=0.45\textwidth]{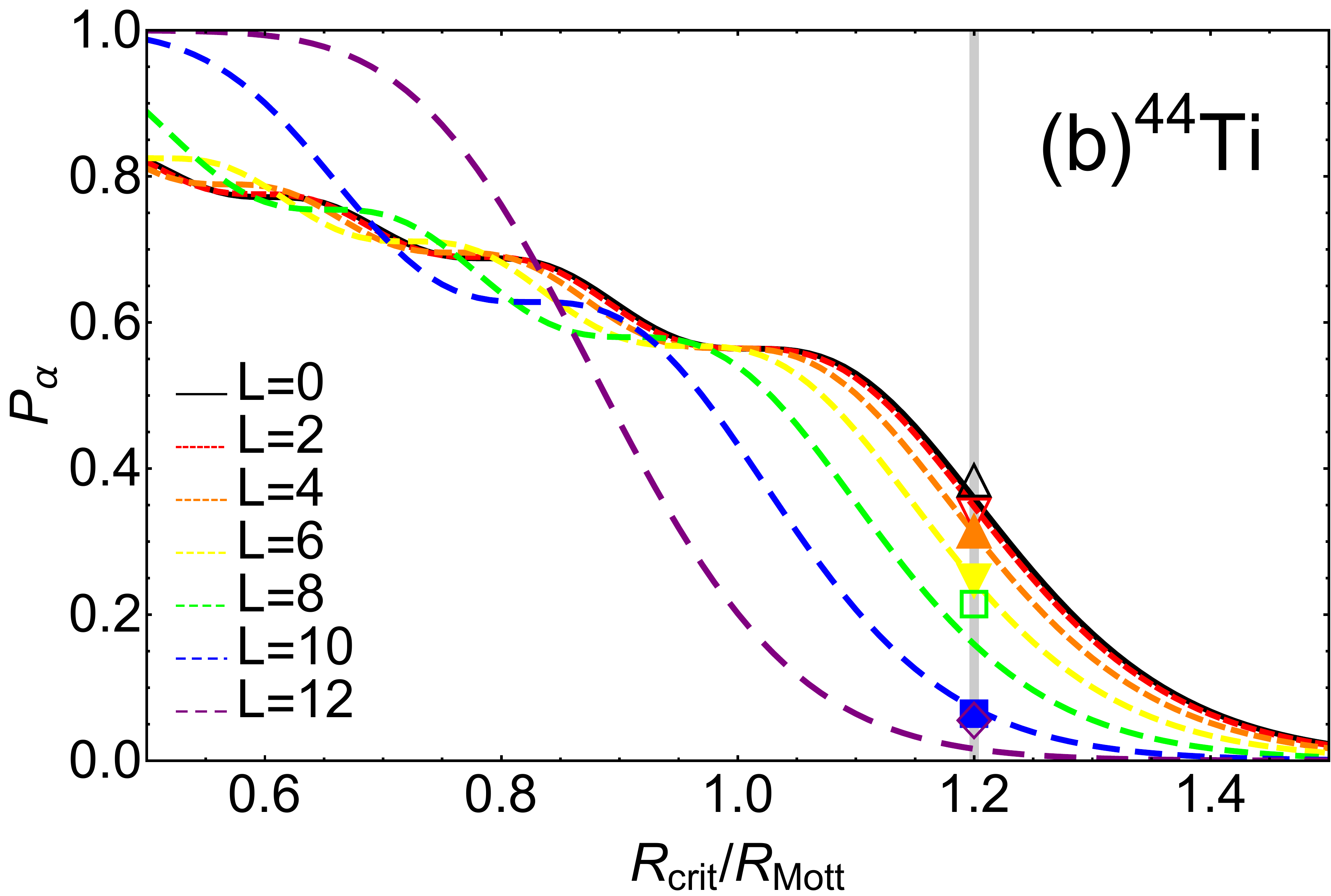}
\end{subfigure}


\begin{subfigure}[b]{\textwidth}
\centering
\includegraphics[width=0.45\textwidth]{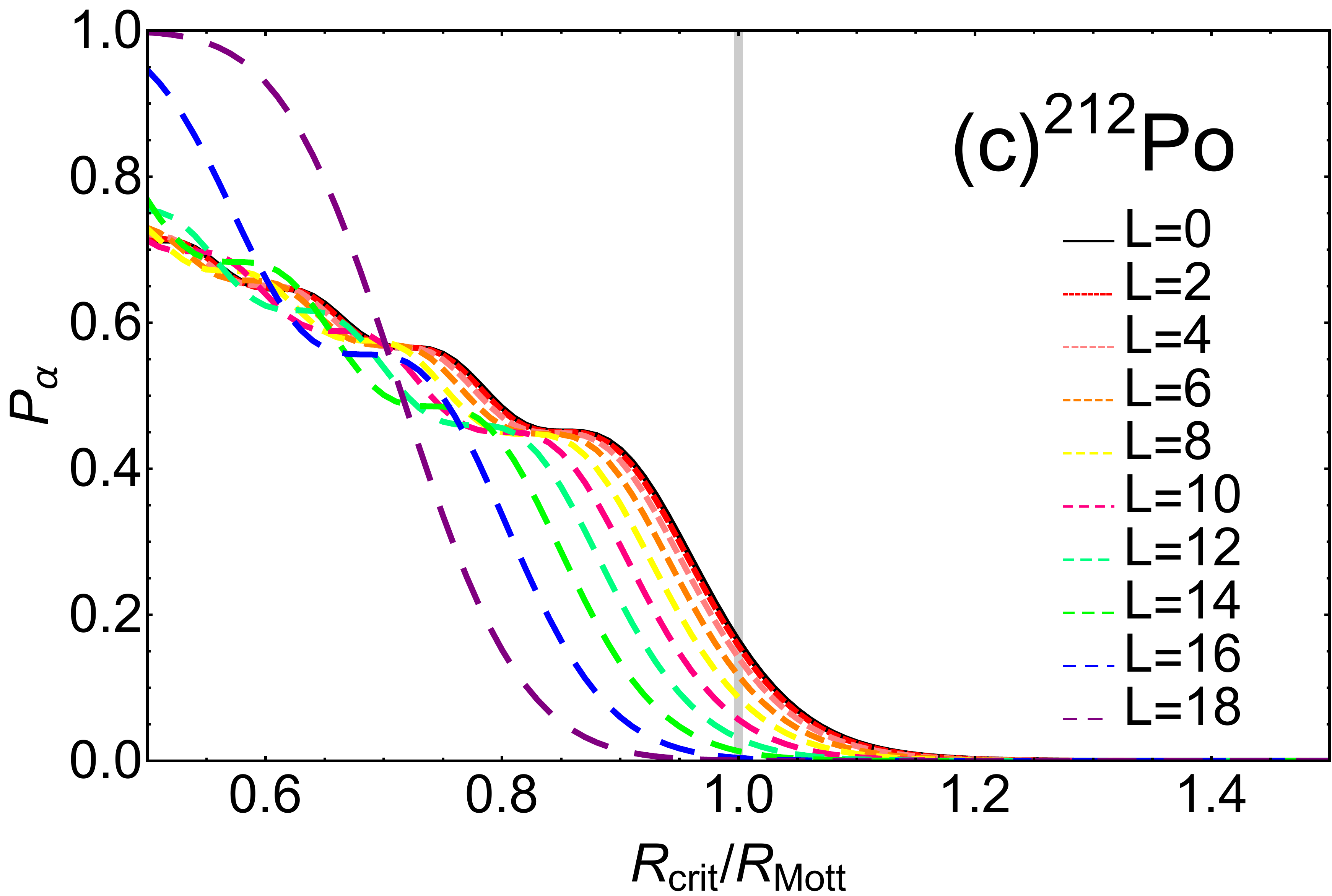}
\end{subfigure}

\caption{The alpha-cluster formation probability $P_\alpha$ vs the critical radius $R_\text{crit}$} for the ground-state bands of $^{20}$Ne, $^{44}$Ti, and $^{212}$Po. In Fig.~\ref{PA}a, the black solid line corresponds to the results for the $L=0$ state, while the dashed lines with increasing segment lengths correspond to the results for $L=2-8$, respectively. The data points in Fig.~\ref{PA}a denote the AMD results on $P_\alpha$ taken from Ref.~\cite{Kimura:2003uf}, with the $L=0$ data point labeled by the empty up triangle, the $L=2$ data point labeled by the empty down triangle, the $L=4$ data point labeled by the filled up triangle,  the $L=6$ data point labeled by the filled down triangle, and the $L=8$ data point labeled by the empty square. In Fig.~\ref{PA}b, the black solid line corresponds to the results for the $L=0$ state, while the dashed lines with increasing segment lengths correspond to the results for $L=2-12$, respectively. The data points \ref{PA}b denote the AMD results on $P_\alpha$ taken from Ref.~\cite{Kimura:2004ez}, with the $L=0$ data point labeled by the empty up triangle, the $L=2$ data point labeled by the empty down triangle, the $L=4$ data point labeled by the filled up triangle, the $L=6$ data point labeled by the filled down triangle, the $L=8$ data point labeled by the empty square, and the $L=10$ data point labeled by the empty diamond. In Fig.~\ref{PA}c, the black solid line corresponds to the results for the $L=0$ state, while the dashed lines with increasing segment lengths correspond to the results for $L=2-18$, respectively.
\label{PA}
\end{figure}

Given the effective potential $W(r)$, the level scheme and the CM wave functions of the quartet structure could be obtained by solving the Schr\"odinger equation Eq.~\eqref{LSE} numerically. { Rigorously speaking, the alpha-cluster states above the disintegration thresholds are Gamow resonances \cite{Gamow:1928,Condon:1928} (see also Ref.~\cite{Delion:2010} for a pedagogic introduction), and the quasibound-state approximation is adopted in the calculation of various structural properties, such as the energy spectrum, the alpha-cluster formation probability, etc.} To get further information on the alpha-cluster formation probability, the critical radius $R_\text{crit}$ which is considered as an empirical parameter has to be fixed. In the above, we have decided the Mott radius for ${}^{20}$Ne, ${}^{44}$Ti, and ${}^{212}$Po as the benchmark value of the critical radius. Here, we propose to use some modified values for $R_\text{crit}$ in the real calculations. Suppose $f_L(r)$ is the radial component of the CM quartet wave function $\varphi_L({\bf r})=f_L( r)/r Y_{LM}(\vartheta, \phi)$ with the angular momentum $L$ that is normalized by $\int\mathrm{d}r|f_L(r)|^2=1$. The alpha-cluster formation probability could be obtained by $P_\alpha(R_\text{crit})=\int_{R_\text{crit}}^{\infty}\mathrm{d}r\left|f_L(r)\right|^2$. In Fig.~\ref{PA}, we consider the relation between the alpha-cluster formation probability $P_\alpha$ and the critical radius $R_\text{crit}$ for the ground and excited states of ${}^{20}$Ne, ${}^{44}$Ti, and ${}^{212}$Po, where the data points correspond to the antisymmetrized molecular dynamics (AMD) results for ${}^{20}$Ne and ${}^{44}$Ti \cite{Kimura:2003uf,Kimura:2004ez}. The AMD approach is a microscopic framework for nuclear cluster physics, which treats the dynamics of nucleons without making any presumption on the existence of cluster structures \cite{Ono:1991uz,KanadaEnyo:1995tb,Ono:1992uy}. By adapting to the AMD results, the critical radius for ${}^{20}$Ne and ${}^{44}$Ti are determined to be $R_\text{crit}=1.2R_\text{Mott}$. The deviation from the Mott radius of the homogeneous nuclear matter is not unexpected as their mass numbers are relatively small and the finite-size effects may be large. For ${}^{212}$Po, on the other hand, we take $R_\text{crit}=R_\text{Mott}$, as its mass number is quite large, which makes its physical properties be closer to those of the homogeneous nuclear matter. The numerical results of the alpha-cluster formation probabilities for the ground and excited states with the spin/parity $J^\pi$ could be found in Table \ref{Ne20Table}, \ref{Ti44Table}, and \ref{Po212Table}. The numerical results  are consistent with the AMD results for ${}^{20}$Ne and ${}^{44}$Ti, as well as previous estimations on the ground-state alpha-cluster formation probability of ${}^{212}$Po \cite{Xu:2015pvv}.


Fig.~\ref{Ne20WF}, \ref{Ti44WF}, and \ref{Po212WF} show explicitly the radial component $f_L(r)$ of the CM wave function $\varphi(r)$ for the ground and excited states of ${}^{20}$Ne, ${}^{44}$Ti, and ${}^{212}$Po, and highlight, in particular, the region where the intrinsic alpha-cluster state is formed. The Wildermuth condition could be checked explicitly by counting the number of the nodes in these radial wave functions for different orbital angular momenta. For instance, the radial wave function for the ground state of ${}^{20}$Ne in Fig.~\ref{Ne20WF}a has four nodes just as required by the Wildermuth condition. These wave functions describe the CM motion of the quartet, and should be distinguished from the CM wave function in the traditional cluster model and the alpha-cluster formation amplitude in the cluster-configuration shell model. The QM approach shows the inner oscillation of the radial wave function, which could be of interest for studying the electromagnetic transitions and nuclear radii. 

Having the CM component of the quartet wave function, we can also calculate the reduced quadrupole transition strength ${B}(\text{E}2\!\downarrow)$ and the rms intercluster separation for the ground and excited states of ${}^{20}$Ne, ${}^{44}$Ti, and ${}^{212}$Po. The numerical results could also be found in Table \ref{Ne20Table}, \ref{Ti44Table}, and \ref{Po212Table}.
The reduced quadrupole transition strength ${B}(\text{E}2\!\downarrow)$ (in the the Weisskopf unit $1 \text{ W.u.}=\frac{0.746}{4\pi}A^{4/3}e^2\cdot \text{fm}^4$ with $A$ being the mass number of the parent nucleus) is obtained by \cite{Buck:1995zza,Bai:2018hbe}
\begin{align}
&{B}(\text{E}2\!\downarrow)=\frac{15\beta_2^2}{8\pi}\frac{L(L-1)}{(2L+1)(2L-1)}\left|\int_0^\infty\mathrm{d}r\, r^2 f_{L-2}(r)^* f_L(r)\right|^2,\\
&\beta_2=e\frac{Z_cA_\alpha^2+Z_\alpha A_c^2}{(A_\alpha+A_c)^2}.
\end{align} 
Here, $A_\alpha$, $Z_\alpha$, $A_c$, and $Z_c$ are the mass numbers and the charge numbers for the alpha cluster and the core nucleus, respectively. The numerical values of ${B}(\text{E}2\!\downarrow)$ for the WSG potential have been reported in Ref.~\cite{Bai:2018hbe}, and are reproduced here for completeness. The root-mean-square (rms) intercluster separation $R_i$ (in the unit of fm) is obtained by
$R_i=\sqrt{\int_0^\infty\mathrm{d}r\, r^2 |f_L(r)|^2}$.
 It is found that the QM results agree well with the experimental values.
 These numerical results, along with the results of the energy spectrum and the alpha-cluster formation probability, provide evidence for the usefulness of the QM approach. {

\begin{figure}
\centering
\begin{subfigure}[b]{\textwidth}
\centering
\minipage{0.45\textwidth}
  \includegraphics[width=\linewidth]{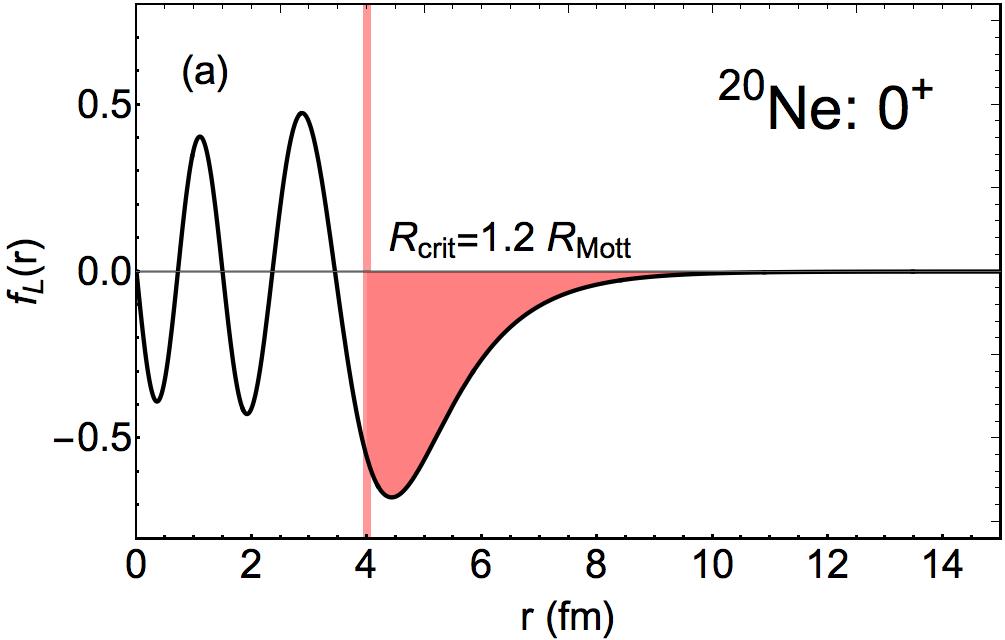}
\endminipage\hfill
\minipage{0.45\textwidth}
  \includegraphics[width=\linewidth]{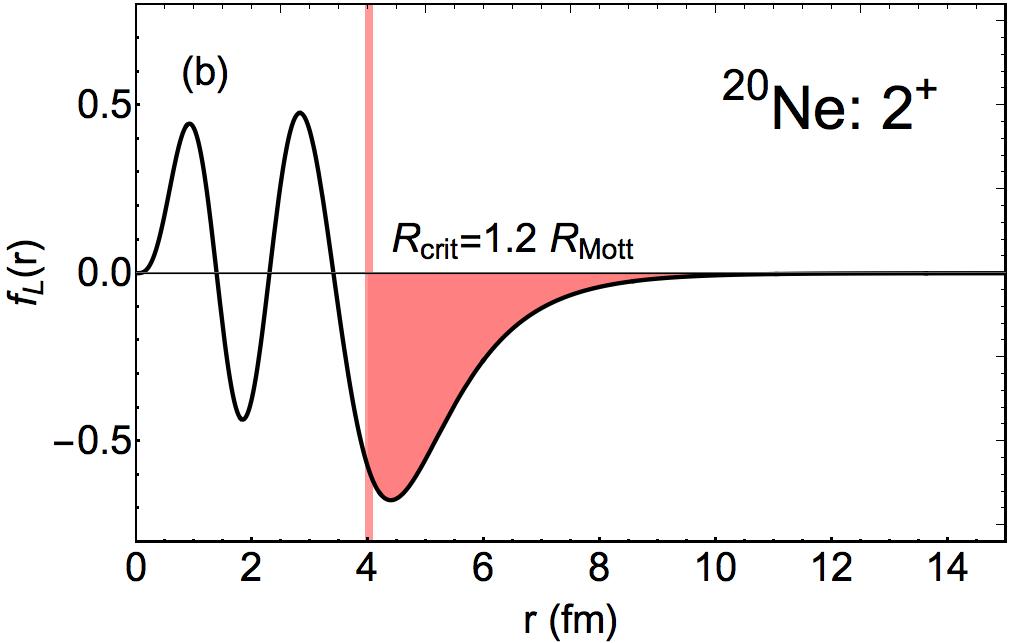}
\endminipage
\end{subfigure}

\begin{subfigure}[b]{\textwidth}
\centering
\minipage{0.45\textwidth}
  \includegraphics[width=\linewidth]{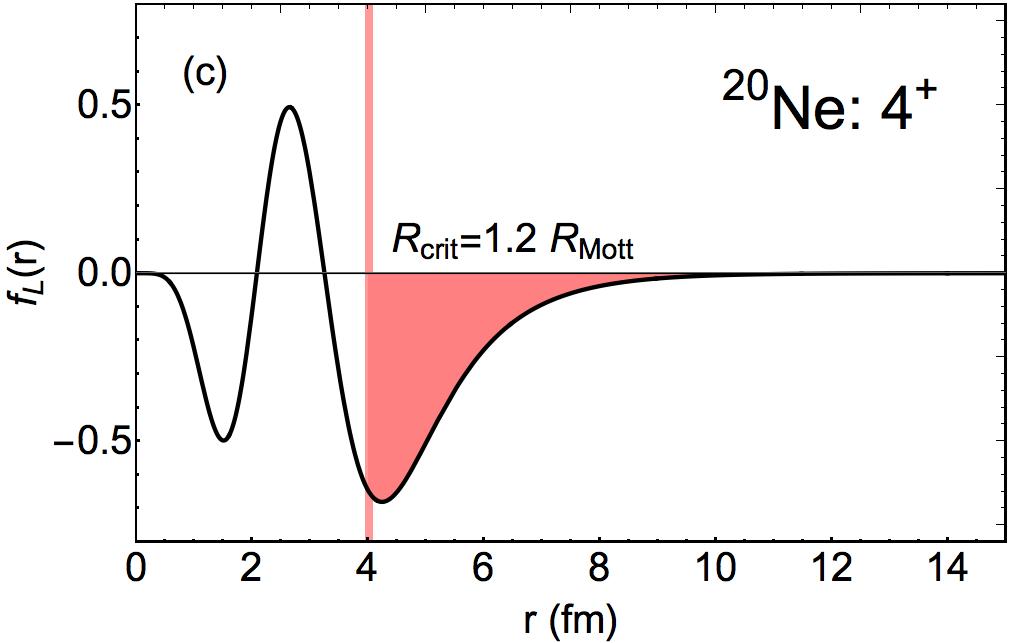}
\endminipage\hfill
\minipage{0.45\textwidth}
  \includegraphics[width=\linewidth]{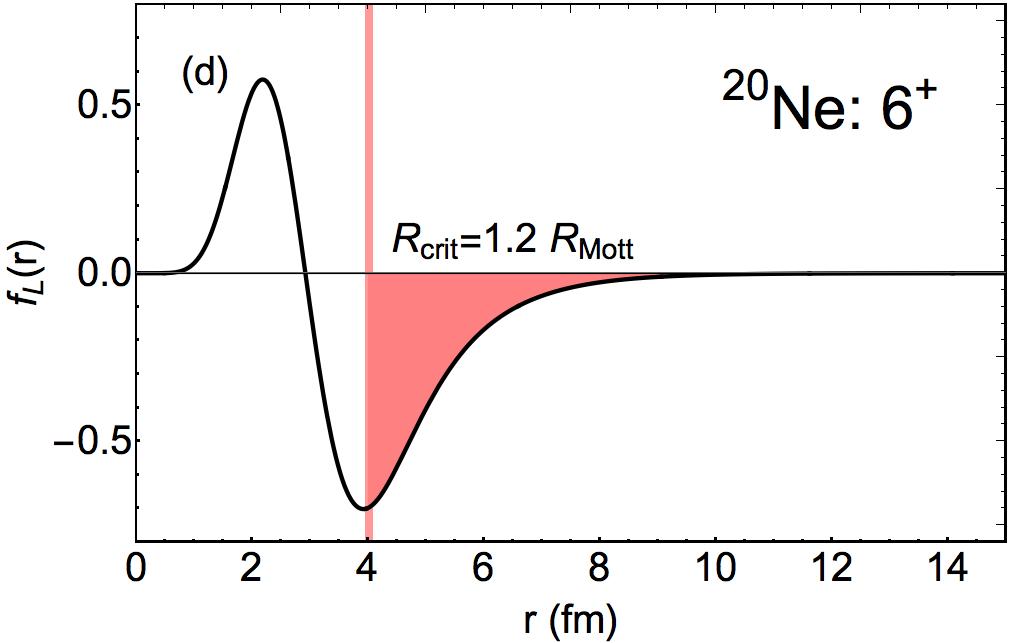}
  \endminipage
\end{subfigure}

\begin{subfigure}[b]{\textwidth}
\centering
\minipage{0.45\textwidth}
  \includegraphics[width=\linewidth]{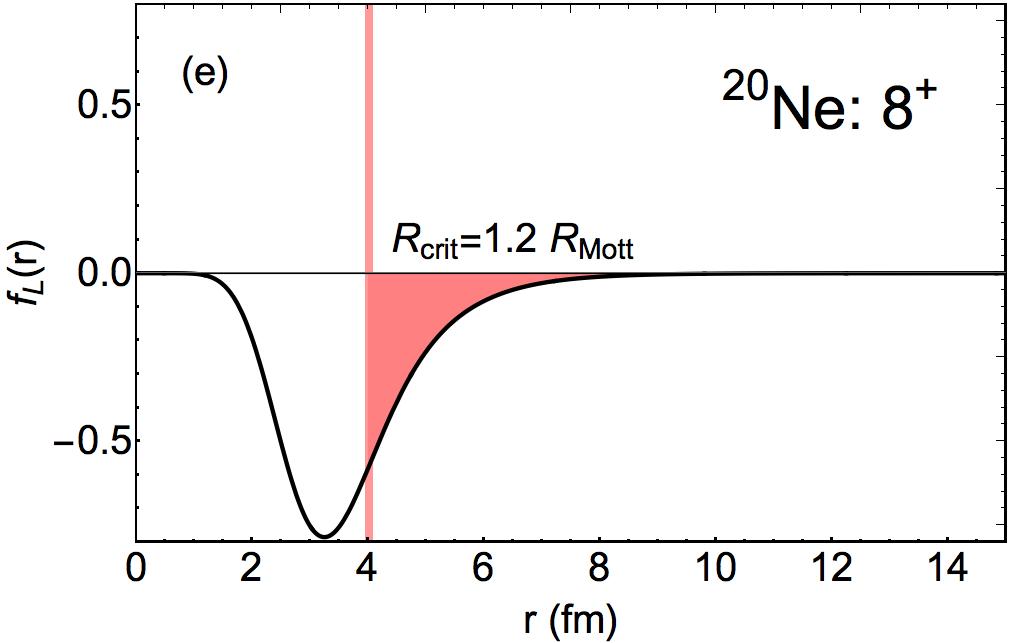}
\endminipage
\end{subfigure}

\caption{The radial components of the quartet wave functions for the ground-state band of $^{20}$Ne. 
The critical radius is taken to be $R_\text{crit}=1.2 R_\text{Mott}$. 
}
\label{Ne20WF}
\end{figure}

\begin{table}
\caption{The QM results for $^{20}$Ne on the energy spectrum, the reduced quadrupole transition strength, the rms intercluster separation, and the alpha-cluster formation probability, along with the experimental values and AMD results for comparison. The experimental values of the energy spectrum and electromagnetic transitions are taken from Ref.~\cite{NNDC,Abele:1993zz}. The AMD values of the alpha-cluster formation probabilities are taken from Ref.~\cite{Kimura:2003uf}.}
\label{Ne20Table}
\begin{center}
\begin{tabular}{ccccccccccc}
\hline
\hline
\hspace{1mm}$J^\pi$\hspace{1mm} & \hspace{1mm}{$E_{\text{exp}}$}\hspace{1mm} & \hspace{1mm}{$E_{\text{th}}$}\hspace{1mm} &  \hspace{1mm}{${B(\text{E}2\!\downarrow)_\text{exp}}$}\hspace{1mm} & \hspace{1mm}{${B(\text{E}2\!\downarrow)_\text{th}}$}\hspace{1mm} & \hspace{1mm}{$R_i$}\hspace{1mm} & \hspace{1mm}{$P_\alpha(\text{AMD})$}\hspace{1mm}  &  \hspace{1mm}{$P_\alpha(\text{QM})$}\hspace{1mm}  \\[-2ex]  
\hspace{1mm}{}\hspace{1mm} & \hspace{1mm}{$\text{[MeV]}$}\hspace{1mm} & \hspace{1mm}{$\text{[MeV]}$}\hspace{1mm} &  \hspace{1mm}{$\text{[W.u.]}$}\hspace{1mm} & \hspace{1mm}{$\text{[W.u.]}$}\hspace{1mm} & \hspace{1mm}{$\text{[fm]}$}\hspace{1mm} & \hspace{1mm}{}\hspace{1mm}  &  \hspace{1mm}{}\hspace{1mm}  \\[0.5ex]  
\hline
\hspace{4mm}$0^+$\hspace{4mm} & \hspace{4mm}{0.000}\hspace{4mm} &  \hspace{4mm}{1.196}\hspace{4mm}  & \hspace{4mm}{$-$}\hspace{4mm} & \hspace{4mm}{$-$}\hspace{4mm} & \hspace{4mm}{$4.14$}\hspace{4mm} & \hspace{4mm}{$0.70$}\hspace{4mm} & \hspace{4mm}{$0.61$}\hspace{4mm} \\
\hspace{4mm}$2^+$\hspace{4mm} & \hspace{4mm}{$1.634$}\hspace{4mm} &  \hspace{4mm}{2.320}\hspace{4mm}  & \hspace{4mm}{$20.3\pm1.0$}\hspace{4mm} & \hspace{4mm}{18.3}\hspace{4mm} & \hspace{4mm}{$4.13$}\hspace{4mm} & \hspace{4mm}{$0.68$}\hspace{4mm} & \hspace{4mm}{$0.60$}\hspace{4mm} \\
\hspace{4mm}$4^+$\hspace{4mm} &  \hspace{4mm}{$4.248$}\hspace{4mm} & \hspace{4mm}{4.529}\hspace{4mm}  & \hspace{4mm}{$22.0\pm2.0$}\hspace{4mm} & \hspace{4mm}{23.7}\hspace{4mm} & \hspace{4mm}{$4.04$}\hspace{4mm} & \hspace{4mm}{$0.54$}\hspace{4mm} & \hspace{4mm}{$0.55$}\hspace{4mm} \\
\hspace{4mm}$6^+$\hspace{4mm} & \hspace{4mm}{8.776}\hspace{4mm} &  \hspace{4mm}{7.706}\hspace{4mm} &  \hspace{4mm}{$20.0\pm3.0$}\hspace{4mm} & \hspace{4mm}{19.3}\hspace{4mm} & \hspace{4mm}{$3.83$}\hspace{4mm} & \hspace{4mm}{$0.34$}\hspace{4mm} & \hspace{4mm}{$\!\!\!\!\!\!0.42$}\hspace{4mm}\\
\hspace{4mm}$8^+$\hspace{4mm} &  \hspace{4mm}{11.951}\hspace{4mm} & \hspace{4mm}{11.764}\hspace{4mm}  & \hspace{4mm}{$9.03\pm1.3$}\hspace{4mm} & \hspace{4mm}{9.9}\hspace{4mm} & \hspace{4mm}{$3.50$}\hspace{4mm} & \hspace{4mm}{$0.28$}\hspace{4mm} & \hspace{4mm}{$\!\!\!\!\!\!0.18$}\hspace{4mm}\\
 \hline
\hline
\end{tabular}
\end{center}
\end{table}

%

\begin{figure}
\centering
\begin{subfigure}[b]{\textwidth}
\centering
\minipage{0.45\textwidth}
  \includegraphics[width=\linewidth]{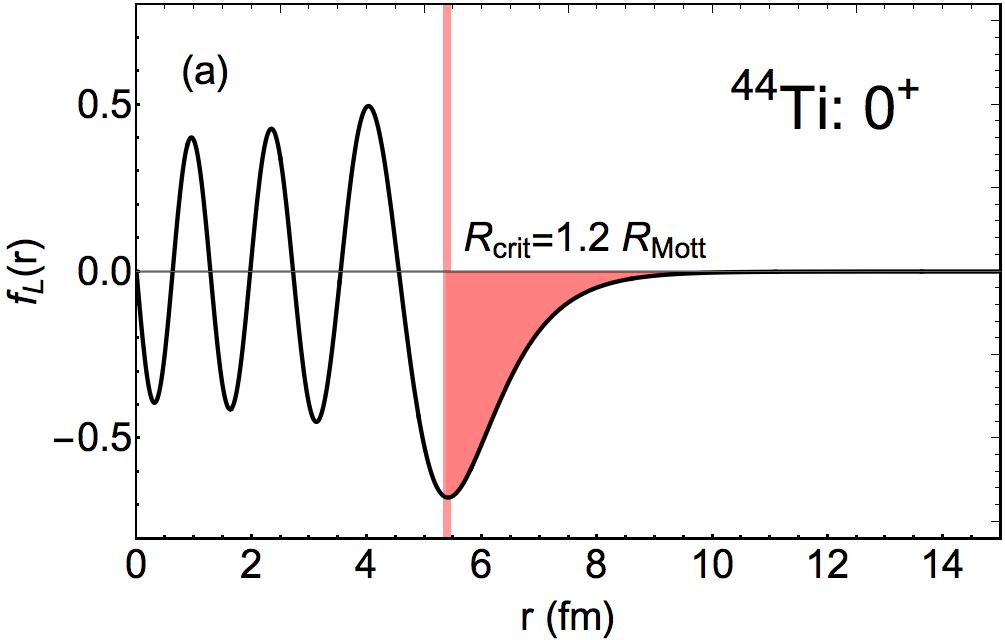}
\endminipage\hfill
\minipage{0.45\textwidth}
  \includegraphics[width=\linewidth]{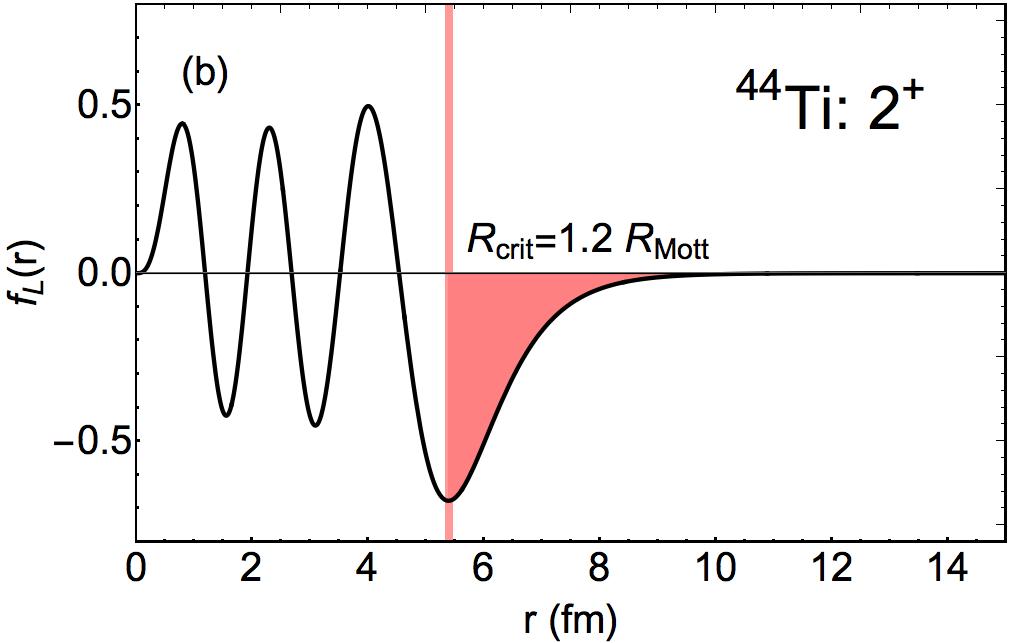}
\endminipage
\end{subfigure}

\begin{subfigure}[b]{\textwidth}
\centering
\minipage{0.45\textwidth}
  \includegraphics[width=\linewidth]{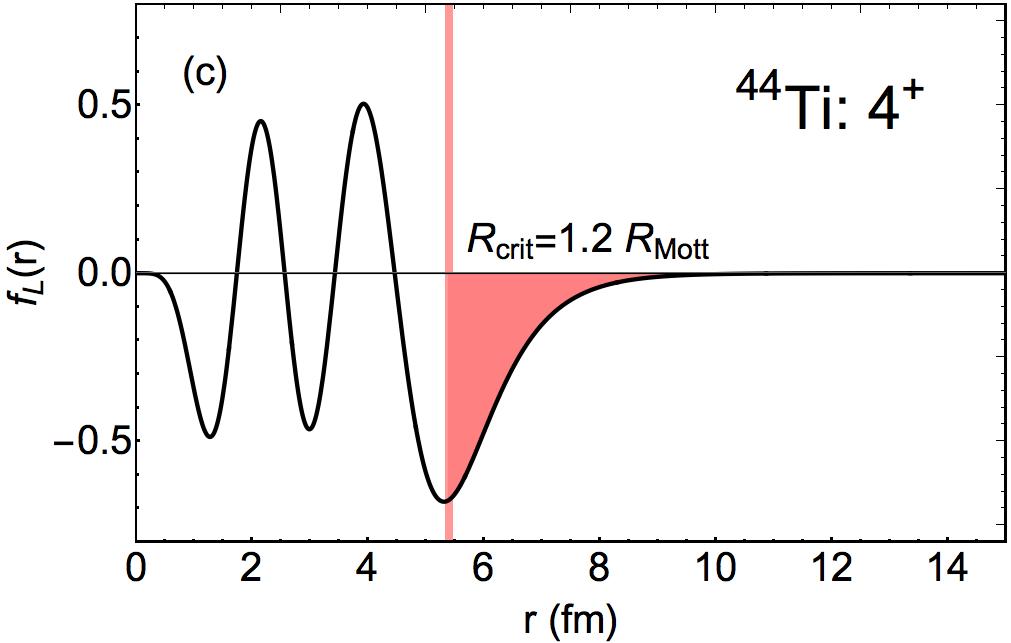}
\endminipage\hfill
\minipage{0.45\textwidth}
  \includegraphics[width=\linewidth]{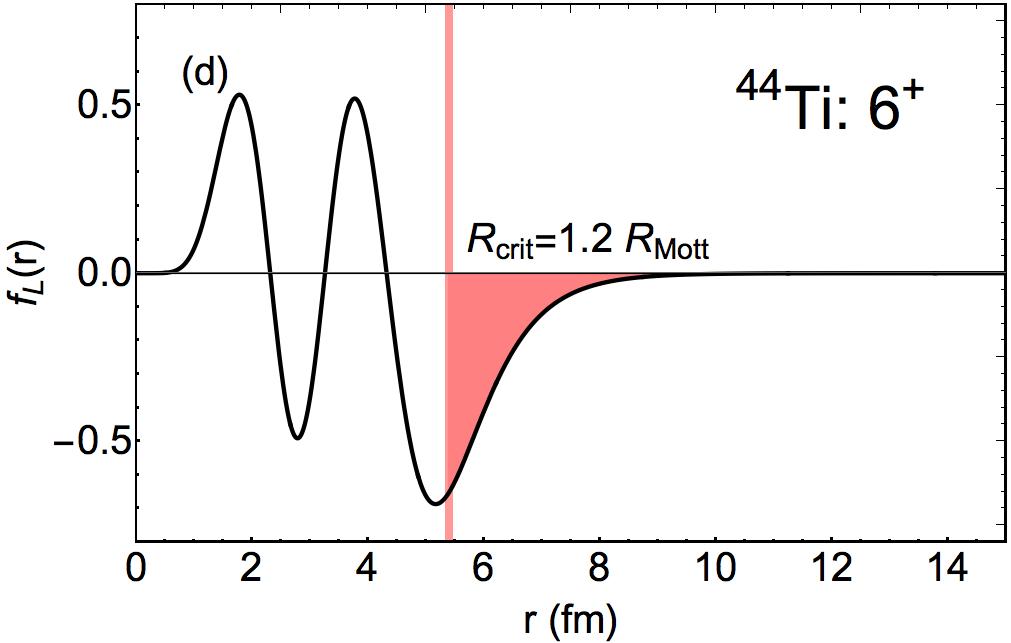}
  \endminipage
\end{subfigure}

\begin{subfigure}[b]{\textwidth}
\centering
\minipage{0.45\textwidth}
  \includegraphics[width=\linewidth]{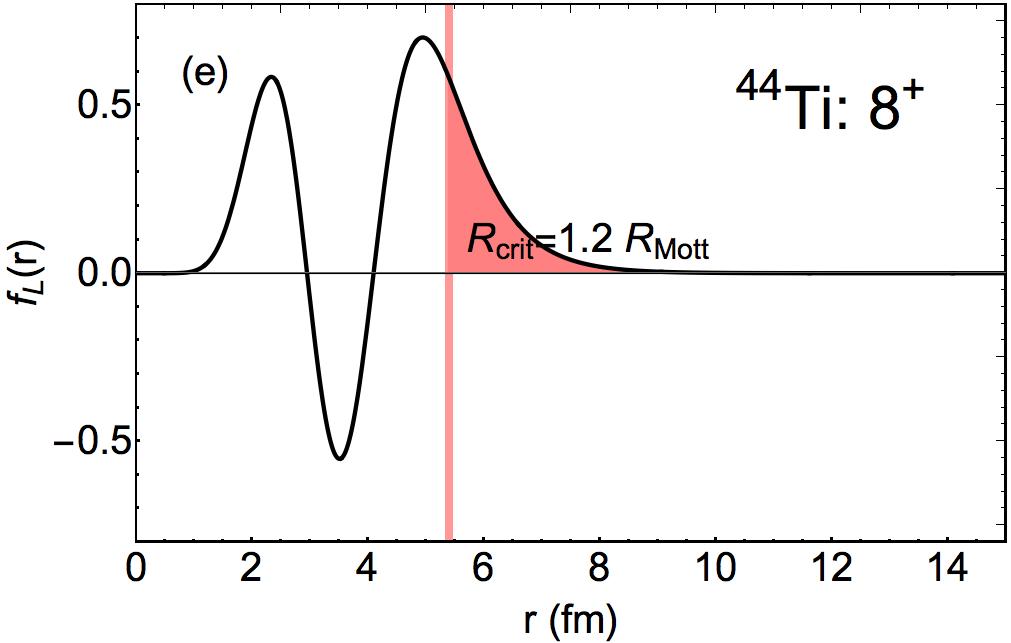}
\endminipage\hfill
\minipage{0.45\textwidth}
  \includegraphics[width=\linewidth]{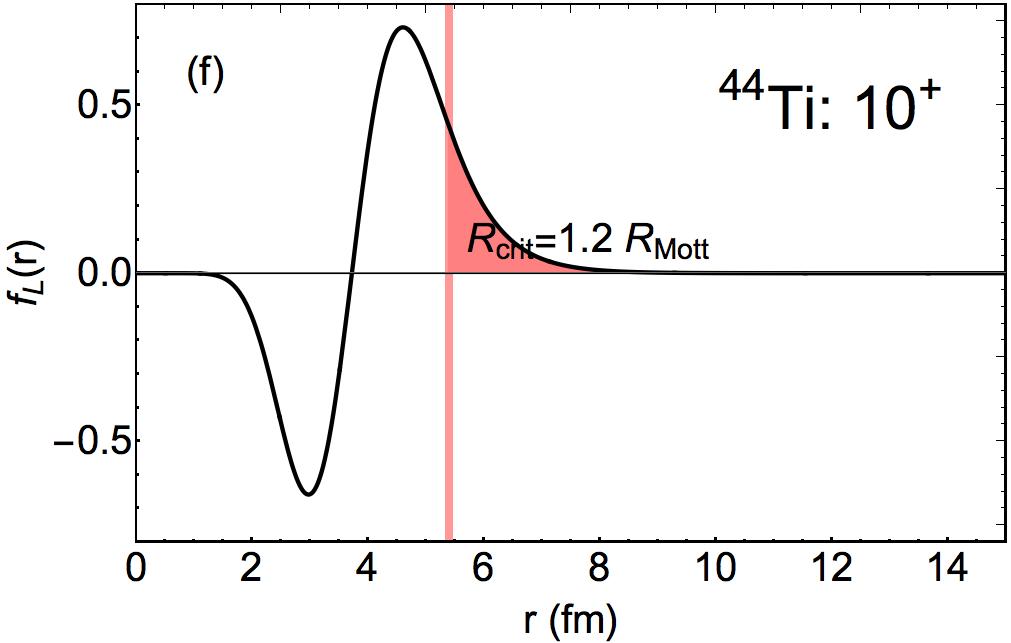}
  \endminipage
\end{subfigure}

\begin{subfigure}[b]{\textwidth}
\centering
\minipage{0.45\textwidth}
  \includegraphics[width=\linewidth]{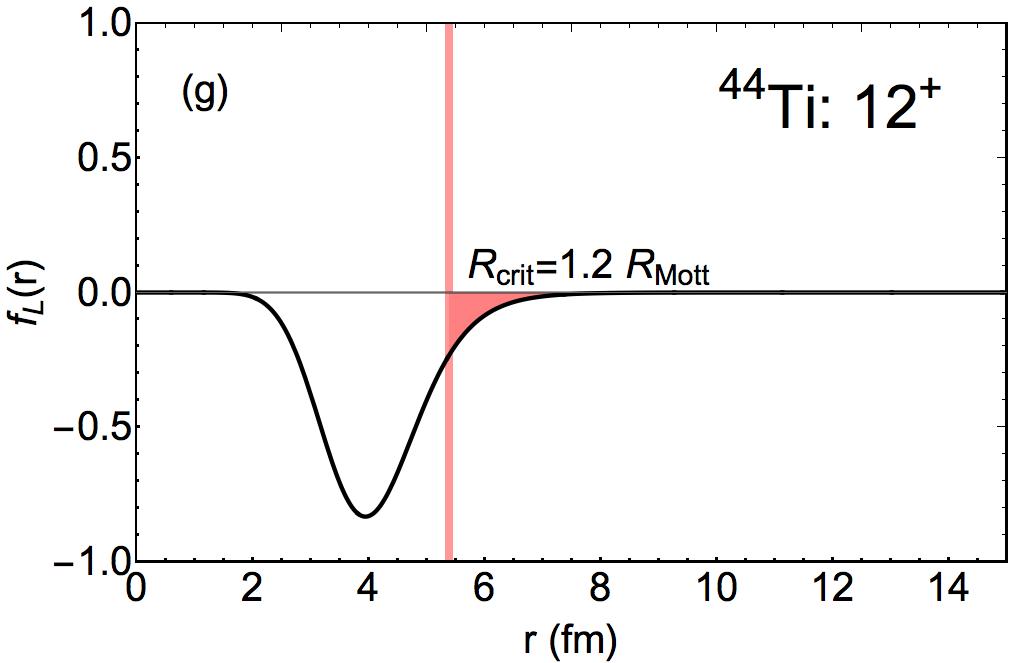}
\endminipage
\end{subfigure}

\caption{The same as Fig.~\ref{Ne20WF}, except that the target nucleus is $^{44}$Ti.}
\label{Ti44WF}
\end{figure}

\begin{table}
\caption{The same as Table \ref{Ne20Table}, except that the target nucleus is $^{44}$Ti. The experimental data are taken from Ref.~\cite{NNDC,Buck:1995zza}. The AMD result is taken from Ref.~\cite{Kimura:2004ez}.}
\label{Ti44Table}
\begin{center}
\begin{tabular}{ccccccccccc}
\hline
\hline
\hspace{1mm}$J^\pi$\hspace{1mm} & \hspace{1mm}{$E_{\text{exp}}$}\hspace{1mm} & \hspace{1mm}{$E_{\text{th}}$}\hspace{1mm} &  \hspace{1mm}{${B(\text{E}2\!\downarrow)_\text{exp}}$}\hspace{1mm} & \hspace{1mm}{${B(\text{E}2\!\downarrow)_\text{th}}$}\hspace{1mm} & \hspace{1mm}{$R_i$}\hspace{1mm} & \hspace{1mm}{$P_\alpha(\text{AMD})$}\hspace{1mm}  &  \hspace{1mm}{$P_\alpha(\text{QM})$}\hspace{1mm}  \\[-2.0ex]  
\hspace{1mm}{}\hspace{1mm} & \hspace{1mm}{$\text{[MeV]}$}\hspace{1mm} & \hspace{1mm}{$\text{[MeV]}$}\hspace{1mm} &  \hspace{1mm}{$\text{[W.u.]}$}\hspace{1mm} & \hspace{1mm}{$\text{[W.u.]}$}\hspace{1mm} & \hspace{1mm}{$\text{[fm]}$}\hspace{1mm} & \hspace{1mm}{}\hspace{1mm}  &  \hspace{1mm}{}\hspace{1mm}  \\[0.5ex]  
\hline
\hspace{4mm}$0^+$\hspace{4mm} & \hspace{4mm}{0.000}\hspace{4mm} &  \hspace{4mm}{0.7673}\hspace{4mm}  & \hspace{4mm}{$-$}\hspace{4mm} & \hspace{4mm}{$-$}\hspace{4mm} & \hspace{4mm}{$4.65$}\hspace{4mm} & \hspace{4mm}{$0.40$}\hspace{4mm} & \hspace{4mm}{$0.36$}\hspace{4mm} \\
\hspace{4mm}$2^+$\hspace{4mm} & \hspace{4mm}{$1.083$}\hspace{4mm} &  \hspace{4mm}{1.349}\hspace{4mm}  & \hspace{4mm}{$13.0\pm4.0$}\hspace{4mm} & \hspace{4mm}{13.2}\hspace{4mm} & \hspace{4mm}{$4.63$}\hspace{4mm} & \hspace{4mm}{$0.36$}\hspace{4mm} & \hspace{4mm}{$0.35$}\hspace{4mm} \\
\hspace{4mm}$4^+$\hspace{4mm} &  \hspace{4mm}{$2.454$}\hspace{4mm} & \hspace{4mm}{2.432}\hspace{4mm}  & \hspace{4mm}{$30.0\pm6.0$}\hspace{4mm} & \hspace{4mm}{17.7}\hspace{4mm} & \hspace{4mm}{$4.58$}\hspace{4mm} & \hspace{4mm}{$0.33$}\hspace{4mm} & \hspace{4mm}{$0.31$}\hspace{4mm} \\
\hspace{4mm}$6^+$\hspace{4mm} & \hspace{4mm}{4.015}\hspace{4mm} &  \hspace{4mm}{3.874}\hspace{4mm} &  \hspace{4mm}{$17.0\pm3.0$}\hspace{4mm} & \hspace{4mm}{16.9}\hspace{4mm} & \hspace{4mm}{$4.48$}\hspace{4mm} & \hspace{4mm}{$0.25$}\hspace{4mm} & \hspace{4mm}{$\!\!\!\!\!\!0.25$}\hspace{4mm}\\
\hspace{4mm}$8^+$\hspace{4mm} &  \hspace{4mm}{6.509}\hspace{4mm} & \hspace{4mm}{5.526}\hspace{4mm}  & \hspace{4mm}{$>1.5$}\hspace{4mm} & \hspace{4mm}{13.8}\hspace{4mm} & \hspace{4mm}{$4.36$}\hspace{4mm} & \hspace{4mm}{$0.21$}\hspace{4mm} & \hspace{4mm}{$\!\!\!\!\!\!0.16$}\hspace{4mm}\\
\hspace{4mm}$10^+$\hspace{4mm} &  \hspace{4mm}{7.671}\hspace{4mm} & \hspace{4mm}{7.178}\hspace{4mm}  & \hspace{4mm}{$15.0\pm3.0$}\hspace{4mm} & \hspace{4mm}{9.3}\hspace{4mm} & \hspace{4mm}{$4.21$}\hspace{4mm} & \hspace{4mm}{$0.06$}\hspace{4mm} & \hspace{4mm}{$\!\!\!\!\!\!0.072$}\hspace{4mm}\\
\hspace{4mm}$12^+$\hspace{4mm} &  \hspace{4mm}{8.039}\hspace{4mm} & \hspace{4mm}{8.528}\hspace{4mm}  & \hspace{4mm}{$<6.5$}\hspace{4mm} & \hspace{4mm}{4.5}\hspace{4mm} & \hspace{4mm}{$4.07$}\hspace{4mm} & \hspace{4mm}{$0.05$}\hspace{4mm} & \hspace{4mm}{$\!\!\!\!\!\!0.016$}\hspace{4mm}\\

 \hline
\hline
\end{tabular}
\end{center}
\end{table}


%

\begin{figure}
\centering

\begin{subfigure}[b]{\textwidth}
\centering
\minipage{0.4\textwidth}
  \includegraphics[width=\linewidth]{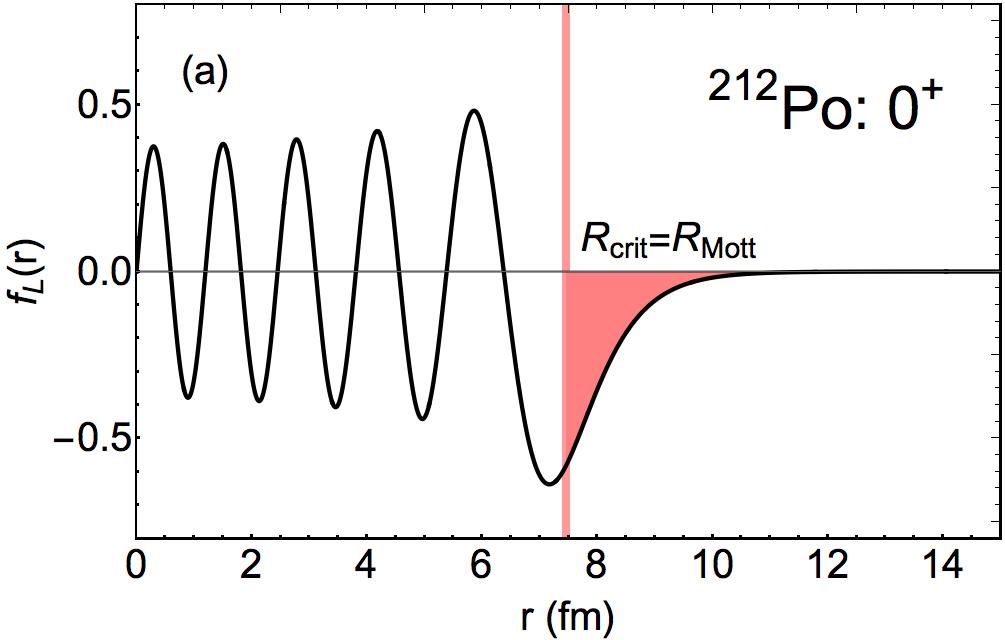}
\endminipage\hfill
\minipage{0.4\textwidth}
  \includegraphics[width=\linewidth]{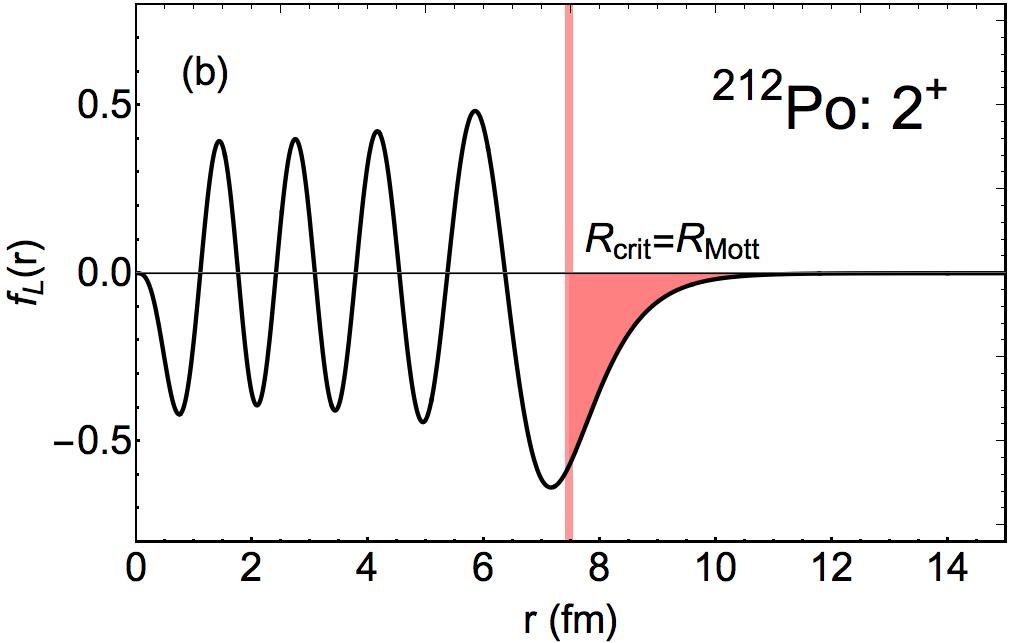}
\endminipage
\end{subfigure}

\begin{subfigure}[b]{\textwidth}
\centering
\minipage{0.4\textwidth}
  \includegraphics[width=\linewidth]{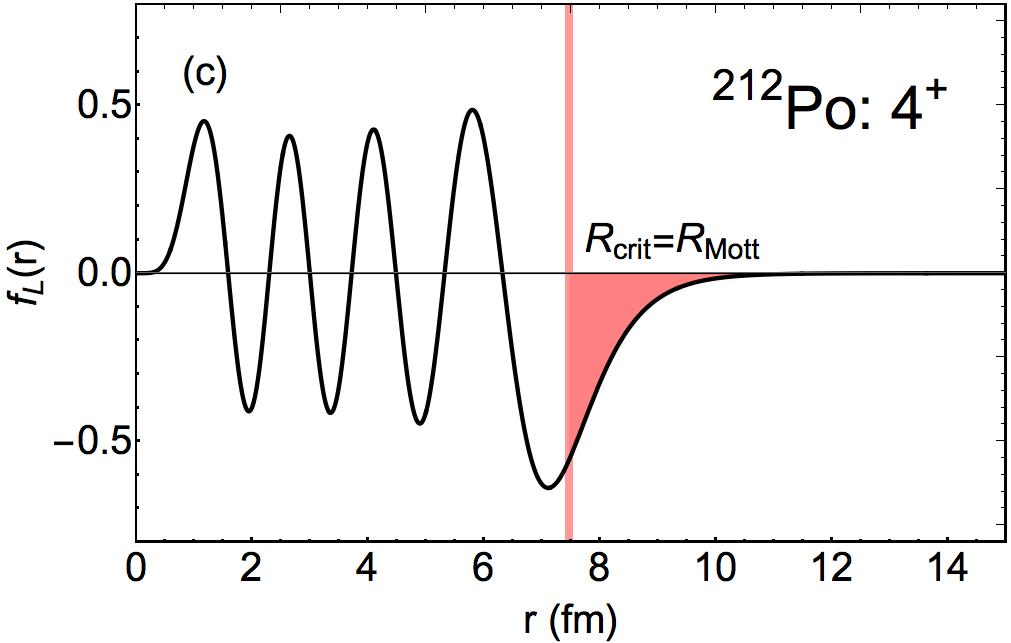}
\endminipage\hfill
\minipage{0.4\textwidth}
  \includegraphics[width=\linewidth]{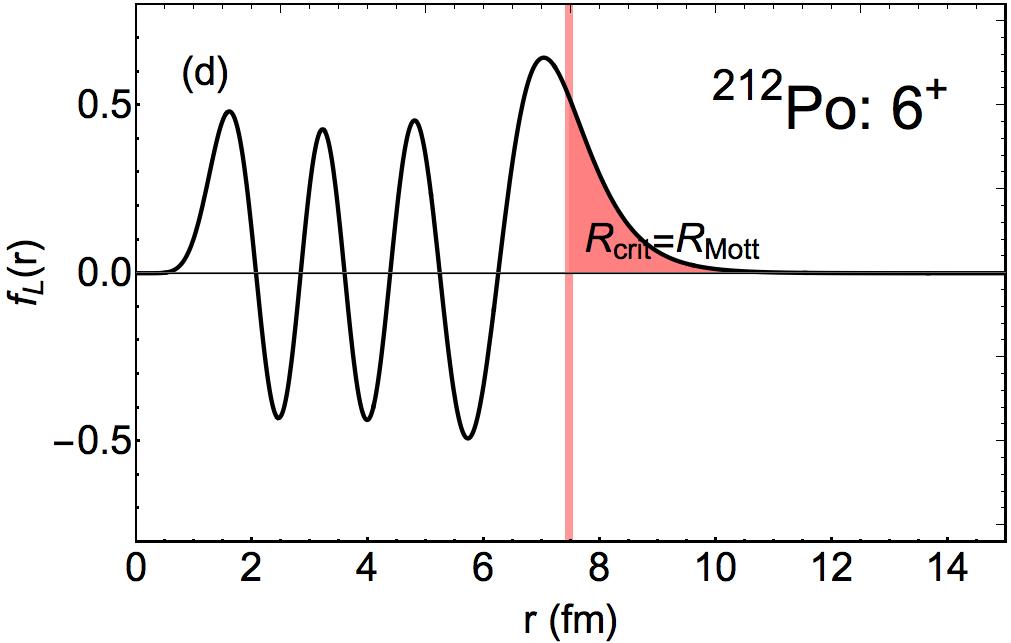}
  \endminipage
\end{subfigure}

\begin{subfigure}[b]{\textwidth}
\centering
\minipage{0.4\textwidth}
  \includegraphics[width=\linewidth]{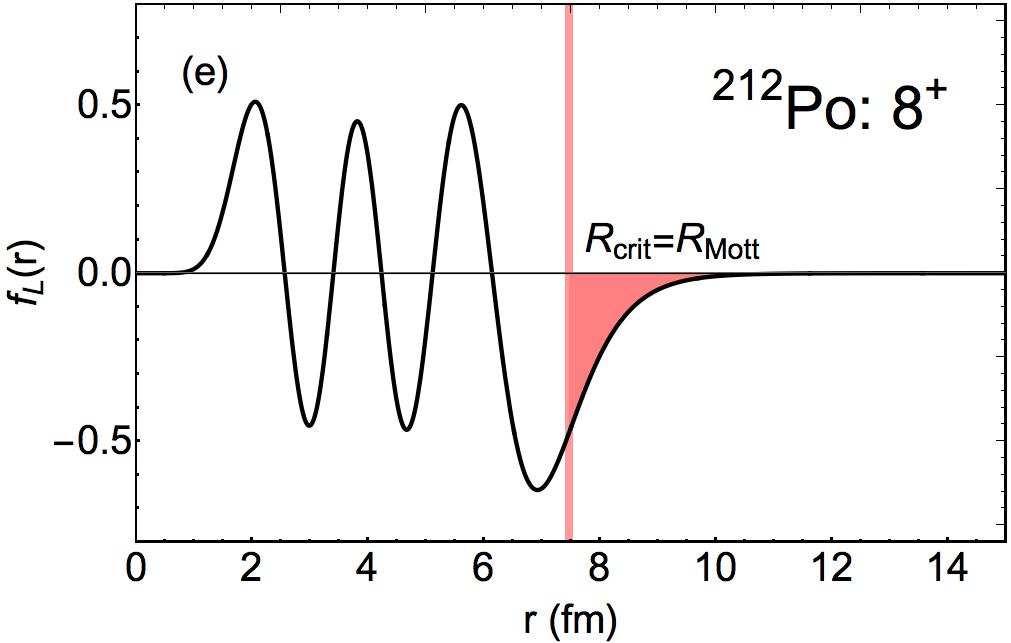}
\endminipage\hfill
\minipage{0.4\textwidth}
  \includegraphics[width=\linewidth]{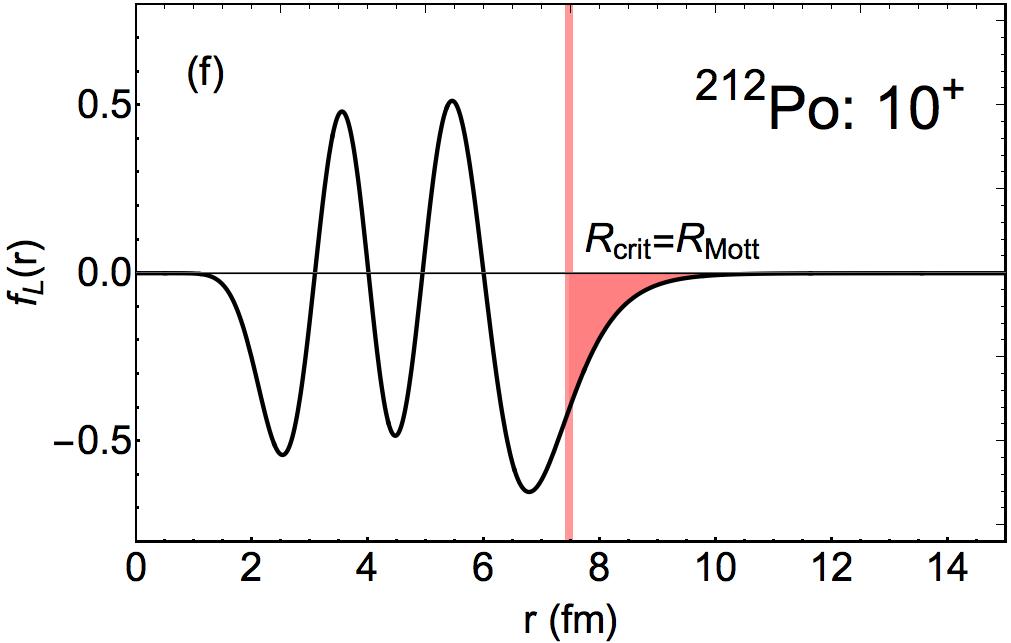}
  \endminipage
\end{subfigure}

\begin{subfigure}[b]{\textwidth}
\centering
\minipage{0.4\textwidth}
  \includegraphics[width=\linewidth]{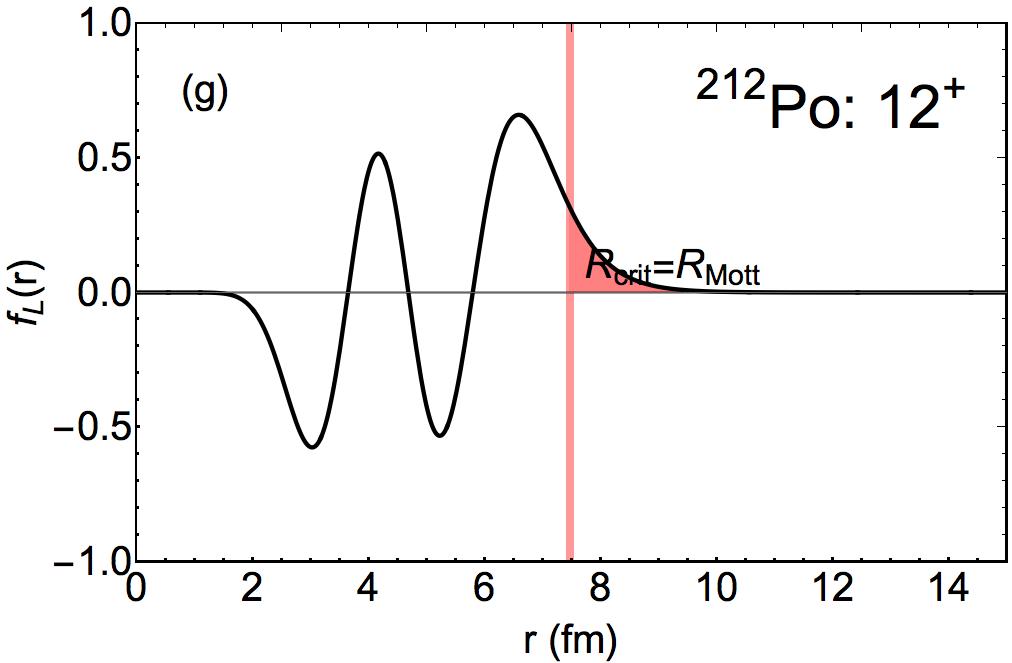}
\endminipage\hfill
\minipage{0.4\textwidth}
  \includegraphics[width=\linewidth]{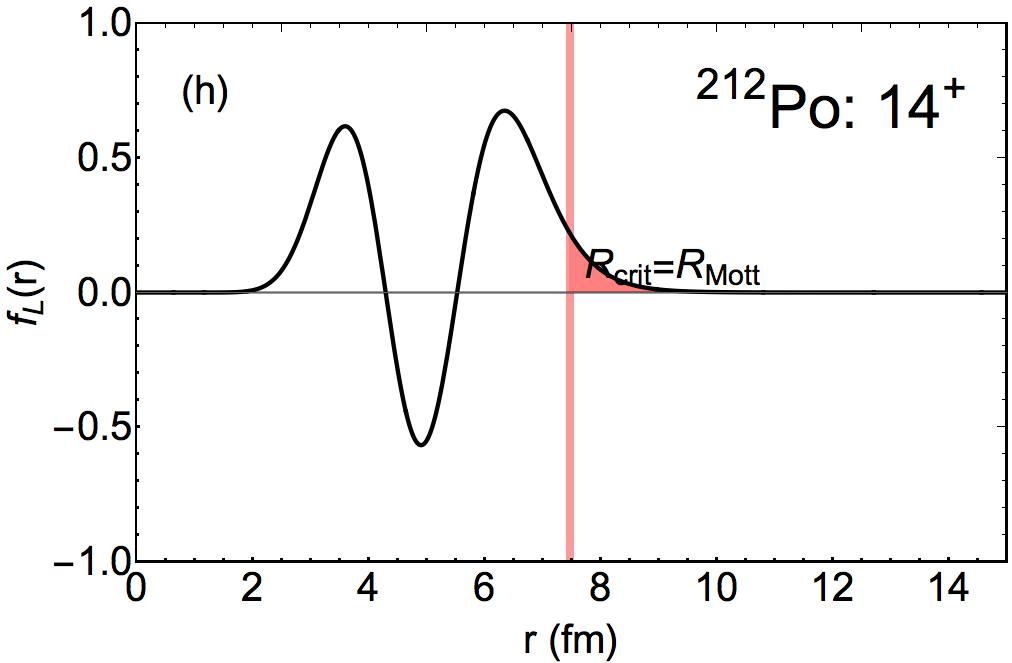}
  \endminipage
\end{subfigure}

\begin{subfigure}[b]{\textwidth}
\centering
\minipage{0.4\textwidth}
  \includegraphics[width=\linewidth]{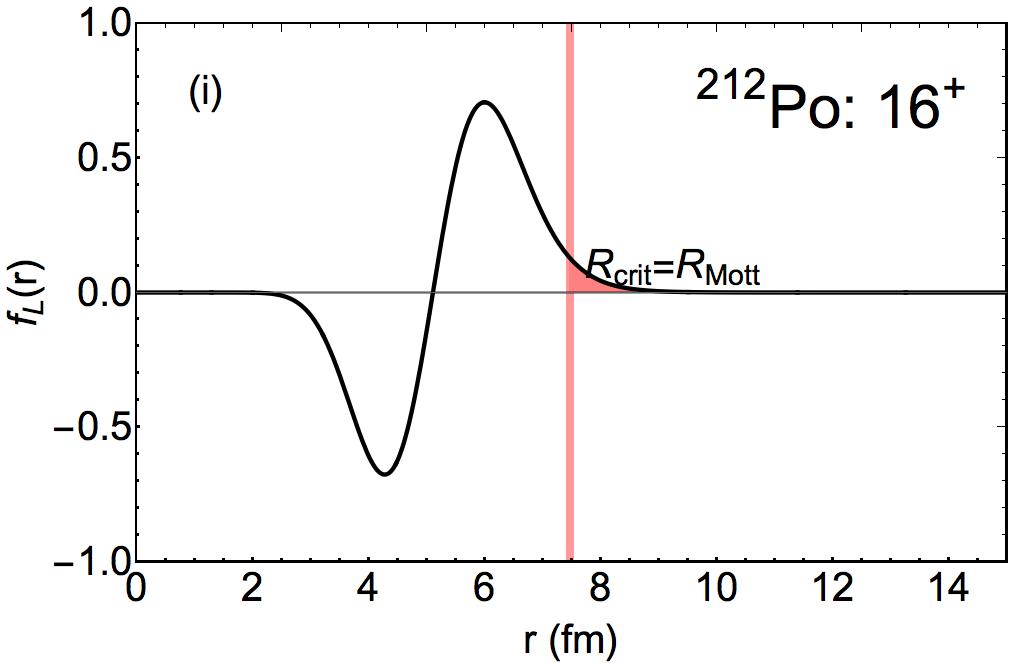}
\endminipage\hfill
\minipage{0.4\textwidth}
  \includegraphics[width=\linewidth]{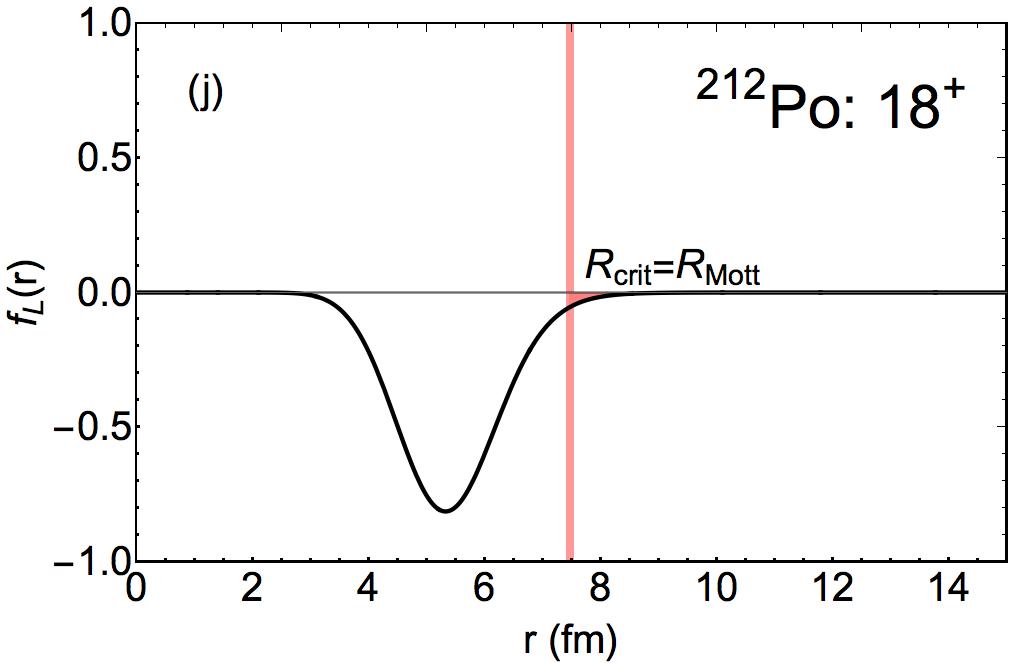}
  \endminipage
\end{subfigure}

\caption{The same as Fig.~\ref{Ne20WF}, except that target nucleus is $^{212}$Po and the critical radius is $R_\text{crit}=1.2R_\text{Mott}$.}
\label{Po212WF}
\end{figure}

\begin{table}
\caption{The same as Table \ref{Ne20Table}, except that the target nucleus is $^{212}$Po. The experimental values are taken from Ref.~\cite{Astier:2009bs,Astier:2010sn}.}
\label{Po212Table}
\begin{center}
\begin{tabular}{cccccccccc}
\hline
\hline
\hspace{2mm}$J^\pi$\hspace{2mm} & \hspace{2mm}{$E_{\text{exp}}$}\hspace{2mm} & \hspace{2mm}{$E_{\text{th}}$}\hspace{2mm} &  \hspace{2mm}{${B(\text{E}2\!\downarrow)_\text{exp}}$}\hspace{2mm} & \hspace{2mm}{${B(\text{E}2\!\downarrow)_\text{th}}$}\hspace{2mm} & \hspace{2mm}{$R_i$}\hspace{2mm}& \hspace{2mm}{$P_\alpha(\text{QM})$}\hspace{2mm} \\[-2.0ex]  
\hspace{1mm}{}\hspace{1mm} & \hspace{1mm}{$\text{[MeV]}$}\hspace{1mm} & \hspace{1mm}{$\text{[MeV]}$}\hspace{1mm} &  \hspace{1mm}{$\text{[W.u.]}$}\hspace{1mm} & \hspace{1mm}{$\text{[W.u.]}$}\hspace{1mm} & \hspace{1mm}{$\text{[fm]}$}\hspace{1mm} & \hspace{1mm}{}\hspace{1mm}  \\[0.5ex]  
\hline
\hspace{4mm}$0^+$\hspace{4mm} & \hspace{4mm}{0.000}\hspace{4mm} &  \hspace{4mm}{-0.072}\hspace{4mm}  & \hspace{4mm}{$-$}\hspace{4mm} & \hspace{4mm}{$-$}\hspace{4mm} & \hspace{4mm}{$5.76$}\hspace{4mm} & \hspace{4mm}{$\ \ \ 0.16$}\hspace{4mm}  \\
\hspace{4mm}$2^+$\hspace{4mm} & \hspace{4mm}{$0.727$}\hspace{4mm} &  \hspace{4mm}{0.111}\hspace{4mm}  & \hspace{4mm}{$-$}\hspace{4mm} & \hspace{4mm}{4.4}\hspace{4mm} & \hspace{4mm}{$5.75$}\hspace{4mm} & \hspace{4mm}{$0.16$}\hspace{4mm}\\
\hspace{4mm}$4^+$\hspace{4mm} &  \hspace{4mm}{$1.132$}\hspace{4mm} & \hspace{4mm}{0.451}\hspace{4mm}  & \hspace{4mm}{$3.9\pm1.1$}\hspace{4mm} & \hspace{4mm}{6.1}\hspace{4mm} & \hspace{4mm}{$5.71$}\hspace{4mm} & \hspace{4mm}{$0.14$}\hspace{4mm}\\
\hspace{4mm}$6^+$\hspace{4mm} & \hspace{4mm}{1.355}\hspace{4mm} &  \hspace{4mm}{0.906}\hspace{4mm} &  \hspace{4mm}{$2.3\pm0.1$}\hspace{4mm} & \hspace{4mm}{6.3}\hspace{4mm} & \hspace{4mm}{$5.67$}\hspace{4mm} & \hspace{4mm}{$0.12$}\hspace{4mm}\\
\hspace{4mm}$8^+$\hspace{4mm} &  \hspace{4mm}{1.476}\hspace{4mm} & \hspace{4mm}{1.439}\hspace{4mm}  & \hspace{4mm}{$2.2\pm0.6$}\hspace{4mm} & \hspace{4mm}{5.9}\hspace{4mm} & \hspace{4mm}{$5.61$}\hspace{4mm} & \hspace{4mm}{$0.09$}\hspace{4mm}\\
\hspace{4mm}$10^+$\hspace{4mm} &  \hspace{4mm}{1.834}\hspace{4mm} & \hspace{4mm}{2.006}\hspace{4mm}  & \hspace{4mm}{$-$}\hspace{4mm} & \hspace{4mm}{5.2}\hspace{4mm} & \hspace{4mm}{$5.54$}\hspace{4mm} & \hspace{4mm}{$0.06$}\hspace{4mm}\\
\hspace{4mm}$12^+$\hspace{4mm} &  \hspace{4mm}{2.702}\hspace{4mm} & \hspace{4mm}{2.550}\hspace{4mm}  & \hspace{4mm}{$-$}\hspace{4mm} & \hspace{4mm}{4.4}\hspace{4mm} & \hspace{4mm}{$5.57$}\hspace{4mm} & \hspace{4mm}{$0.03$}\hspace{4mm}\\
\hspace{4mm}$14^+$\hspace{4mm} &  \hspace{4mm}{2.885}\hspace{4mm} & \hspace{4mm}{2.996}\hspace{4mm}  & \hspace{4mm}{$-$}\hspace{4mm} & \hspace{4mm}{3.4}\hspace{4mm} & \hspace{4mm}{$5.41$}\hspace{4mm} & \hspace{4mm}{$0.01$}\hspace{4mm}\\
\hspace{4mm}$16^+$\hspace{4mm} &  \hspace{4mm}{$-$}\hspace{4mm} & \hspace{4mm}{3.232}\hspace{4mm}  & \hspace{4mm}{$-$}\hspace{4mm} & \hspace{4mm}{2.3}\hspace{4mm} & \hspace{4mm}{$5.38$}\hspace{4mm} & \hspace{4mm}{$0.004$}\hspace{4mm}\\
\hspace{4mm}$18^+$\hspace{4mm} &  \hspace{4mm}{2.921}\hspace{4mm} & \hspace{4mm}{3.089}\hspace{4mm}  & \hspace{4mm}{$-$}\hspace{4mm} & \hspace{4mm}{$-$}\hspace{4mm} & \hspace{4mm}{$5.39$}\hspace{4mm} & \hspace{4mm}{$0.0006$}\hspace{4mm}\\
 \hline
\hline
\end{tabular}
\end{center}
\end{table}



Furthermore, we also investigate the relation between the alpha-core overlap and the alpha-cluster formation probability. In a recent work \cite{Bai:2018}, two of the authors (D.~B.~and Z.~R.) introduce a dimensionless parameter $D=R_i/(R_c+R_\alpha)$ to quantify the degree of the alpha-core overlap, where $R_i$ is the rms intercluster separation introduced above, and $R_c$ and $R_\alpha$ denote the size of the core nucleus and the alpha particle and could be chosen to be their rms point radii. With this parameter, a large (small) alpha-core overlap would correspond to a small (large) $D$ value. It is found that there could be approximately a positive-correlated linear relation between $P_\alpha$ and $D$ in the vicinity of the touching point for the alpha cluster and the core nucleus. The similar analysis is carried out for the ground-state bands of $^{20}$Ne, $^{44}$Ti, and $^{212}$Po within the framework of the QM approach. The results are summarized in Fig.~\ref{CCO}, from which the linear relation between $P_\alpha$ and $D$ could be seen explicitly and could be viewed as another support to Ref.~\cite{Bai:2018}.

\begin{figure}

\centering

\begin{subfigure}[b]{\textwidth}
\centering
\includegraphics[width=0.45\textwidth]{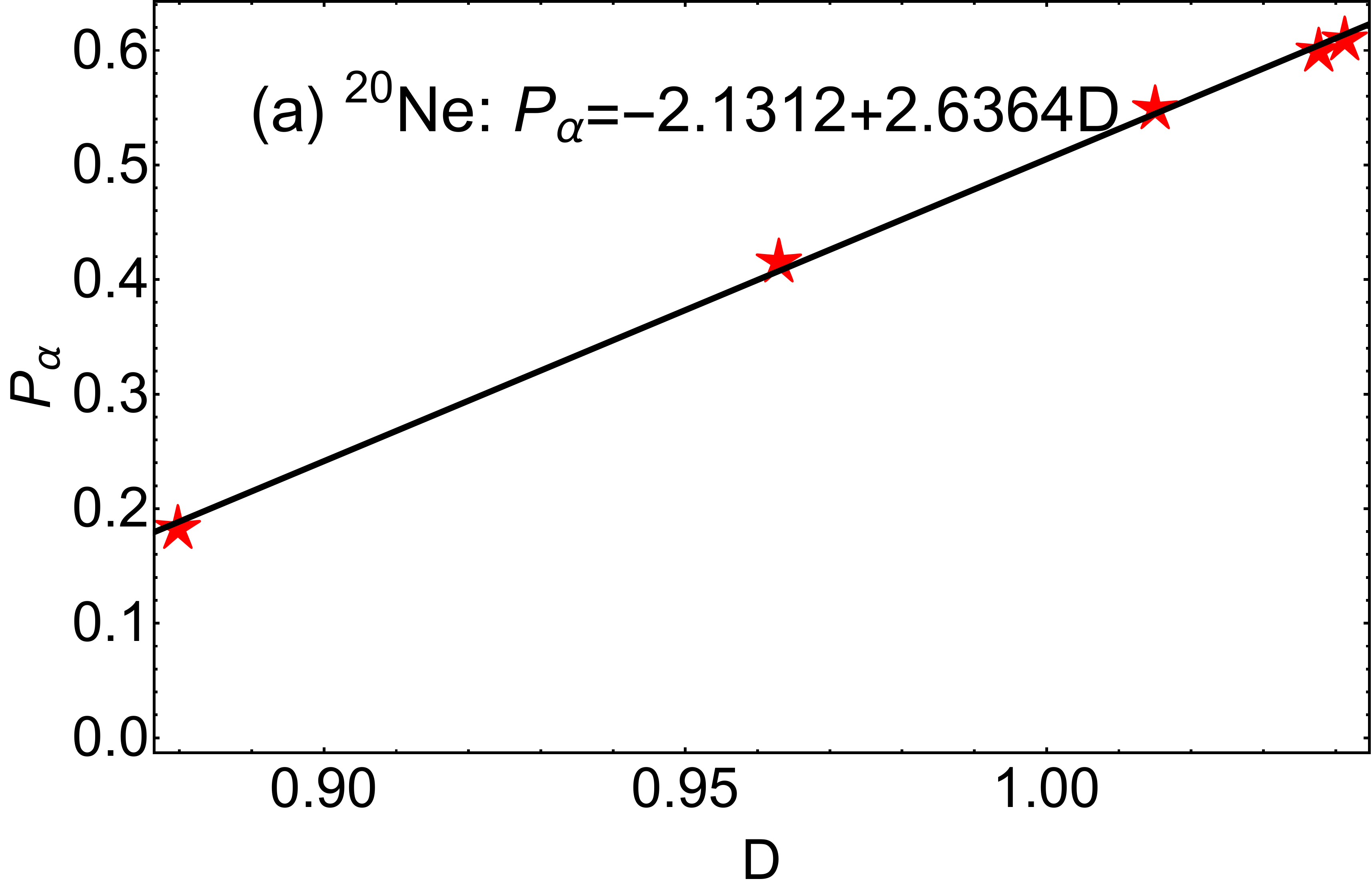}
\end{subfigure}

\begin{subfigure}[b]{\textwidth}
\centering
\includegraphics[width=0.45\textwidth]{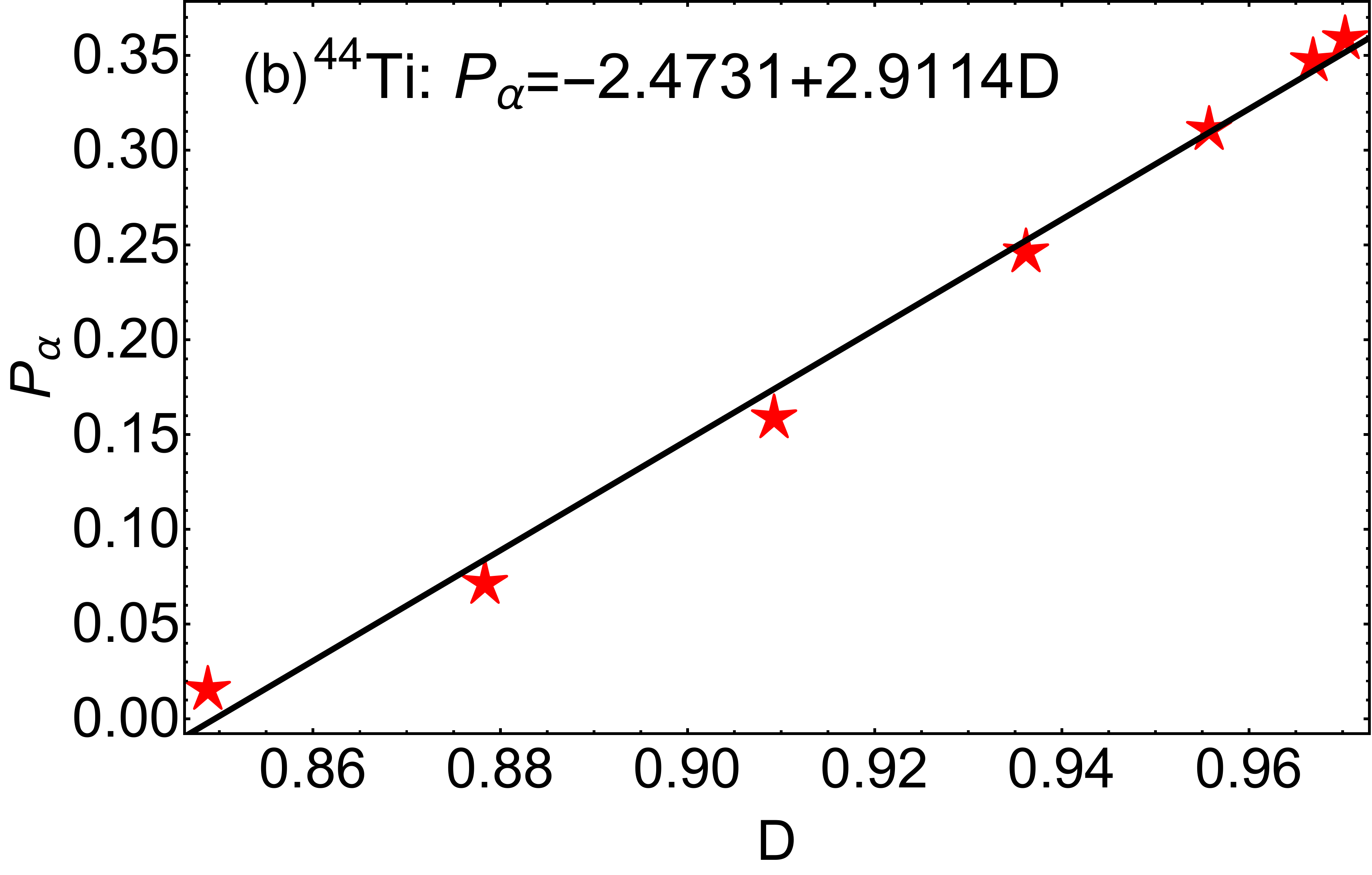}
\end{subfigure}

\begin{subfigure}[b]{\textwidth}
\centering
\includegraphics[width=0.45\textwidth]{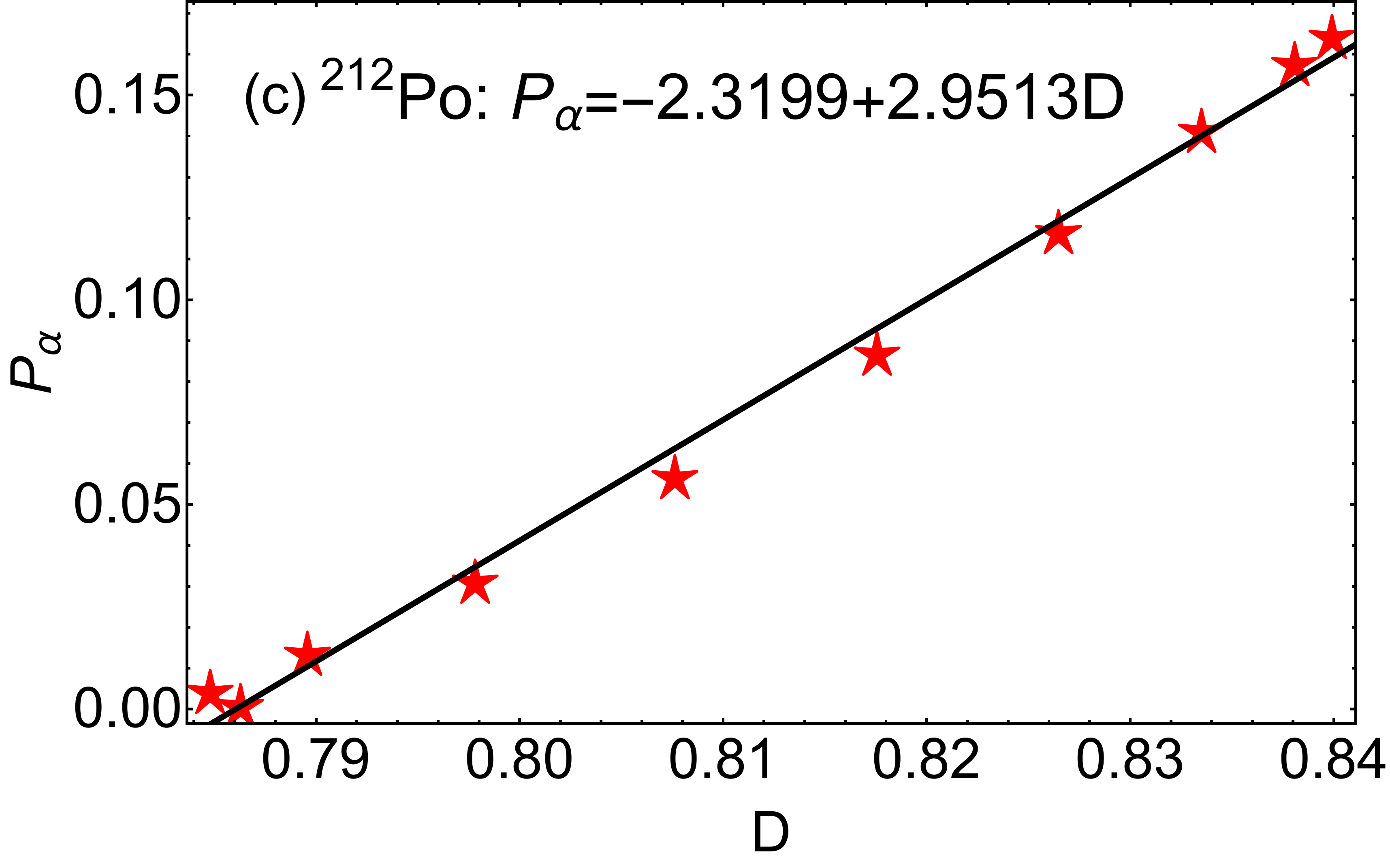}
\end{subfigure}

\caption{The alpha-cluster formation probability vs the alpha-core overlap measured by the parameter $D$ for $^{20}$Ne, $^{44}$Ti, and $^{212}$Po.}
\label{CCO}
\end{figure}

\section{CONCLUSIONS}
\label{Concl}

In this work, we propose the QM approach as a new model to study alpha clustering. It combines various features of both the binary cluster model and the quartetting wave function approach, and aims to provides a unified description of various properties of alpha clustering, including the energy spectrum, electromagnetic transition, nuclear radius, alpha-cluster formation probability, etc. In the QM approach, it is assumed that the intrinsic wave function of the quartet becomes the standard alpha particle when the intercluster separation is larger than the critical radius and merges with the shell-model state of the core nucleus when the intercluster separation is smaller than the critical radius. Then, within the local potential approximation, the relative motion between the quartet and the core nucleus could be solved explicitly. To demonstrate the usefulness of the QM approach, we study the alpha clustering in nuclei where an alpha particle moves on top of a double-magic core nucleus, in particular $^{20}$Ne, $^{44}$Ti, and $^{212}$Po. The effective nuclear potential is chosen to be the WSG potential proposed recently in Ref.~\cite{Bai:2018hbe}, and the effects of the Pauli principle on the relative motion of the quartet and the core nucleus are handled by the Wildermuth condition. The relation to the microscopic quartetting wave function model [9] may be the subject of future investigations. With the WSG parameters and the critical radius chosen properly, the QM approach is shown to be able to give theoretical results that agree well with the experimental data and the previous theoretical results given by AMD simulations. Particularly, we give explicitly the theoretical predictions of the alpha-cluster formation probabilities in the excited states of the $^{212}$Po ground-state band, which could be a useful reference for future studies. Furthermore, we also verify the linear relation between the alpha-cluster formation probability and the alpha-core overlap measure by the parameter $D$ proposed recently in Ref.~\cite{Bai:2018} within the framework of the QM approach. It is well-known that alpha clustering could also manifest itself in more exotic ways like alpha-condensate states in $^{12}$C and other heavier self-conjugate nuclei \cite{Tohsaki:2001an,Yamada:2003cz,Bai:2018gqt,Katsuragi:2018qaj,Barbui:2018sqy}, and it is an important open direction to extend the QM approach to provide a reliable description of them as well. Also, the physical properties of the recent observed new alpha-emitters $^{104}$Te and $^{108}$Xe \cite{Auranen:2018usv,Bai:2018giu} could also be investigated by the QM approach, and this is left for future works. {Furthermore, it is important to make connections between the quartet model and other theoretical models of alpha clustering in literature. For instance, the region above the critical radius could also be obtained in a natural way by using a surface pocket-like potential matched to the Coulomb barrier as shown in Ref.~\cite{Delion:2018whz}.} In summary, we believe that the QM approach could be a valuable complement to the existing models and tools and help deepen our understanding of alpha clustering across the nuclide chart. 

\begin{acknowledgments} 
D.~B.~would like to thank the organizers of the Fourth International Workshop on ``State of the Art in Nuclear Cluster Physics'', May 13-18, 2018, at Galveston, Texas, USA, for the oral talk invitation, where this work is initiated, and the organizers of the Chengdu Workshop on Nuclear Cluster Physics (WNCP2018), November 7-13, 2018, at Chengdu, Sichuan, China, for the oral talk invitation, where this work is discussed. 
This work is supported by the National Key R\&D Program of China (Contract No.~2018YFA0404403, 2016YFE0129300), by the National Natural Science Foundation of China (Grant No.~11535004, 11761161001, 11375086, 11120101005, 11175085, 11235001, 11565010, and 11881240623), and by the Science and Technology Development Fund of Macau under Grant No.~008/2017/AFJ. D.~B.~is also supported by a Project funded by China Postdoctoral Science Foundation (Grant No.~2018M640470).
\end{acknowledgments}

\appendix

\section{DENSITY PROFILES}
\label{DenPro}

In this appendix, we provide the explicit form for the density profiles of the doubly magic nuclei $^{16}$O, $^{40}$Ca, and $^{208}$Pb, which are recompiled from results of Ref.~\cite{DeJager:1987qc} by assuming that the matter density profile is approximately proportional to the charge density profile:
\begin{align}
&\rho_{{}^{16}\text{O}}(r)=0.165362(1-0.00749817 r^2)/\{1+\exp[1.94932(r-2.608)]\}\text{ fm}^{-3},\\
&\rho_{{}^{40}\text{Ca}}(r)=0.169854(1-0.0113518 r^2)/\{1+\exp[1.70648(r-3.766)]\}\text{ fm}^{-3},\\
&\rho_{{}^{208}\text{Pb}}(r)=1.75538\times10^{-6} \exp[{-0.519031 (r-8.7)^2}]+0.00214574 \exp[{-0.519031 (r-7.6)^2}]\nonumber\\
&\qquad\quad\ \ +0.00508279 \exp[{-0.519031 (r-6.6)^2}]+0.0611586 \exp[{-0.519031(r-6)^2}]\nonumber\\
&\qquad\quad\ \ +0.0650727 \exp[{-0.519031 (r-5.1)^2}]+0.0506147 \exp[{-0.519031 (r-4.2)^2}]\nonumber\\
&\qquad\quad\ \ +0.0411758 \exp[{-0.519031 (r-3.5)^2}]+0.0677456 \exp[{-0.519031 (r-2.7)^2}]\nonumber\\
&\qquad\quad\ \ +0.000150248\exp[{-0.519031 (r-2.1)^2}]+0.063191 \exp[{-0.519031 (r-1.6)^2}]\nonumber\\
&\qquad\quad\ \ +0.0450145 \exp[{-0.519031 (r-0.7)^2}]+0.0265771 \exp[{-0.519031 (r-0.1)^2}]\nonumber\\
&\qquad\quad\ \ +0.0265771 \exp[{-0.519031(r+0.1)^2}]+0.0450145 \exp[{-0.519031 (r+0.7)^2}]\nonumber\\
&\qquad\quad\ \ +0.063191 \exp[{-0.519031 (r+1.6)^2}]+0.000150248 \exp[{-0.519031 (r+2.1)^2}]\nonumber\\
&\qquad\quad\ \ +0.0677456 \exp[{-0.519031 (r+2.7)^2}]+0.0411758\exp[{-0.519031 (r+3.5)^2}]\nonumber\\
&\qquad\quad\ \ +0.0506147 \exp[{-0.519031 (r+4.2)^2}]+0.0650727 \exp[{-0.519031 (r+5.1)^2}]\nonumber\\
&\qquad\quad\ \ +0.0611586 \exp[{-0.519031 (r+6)^2}]+0.00508279 \exp[{-0.519031(r+6.6)^2}]\nonumber\\
&\qquad\quad\ \ +0.00214574 \exp[{-0.519031 (r+7.6)^2}]+1.75538\times10^{-6} \exp[{-0.519031 (r+8.7)^2}].
\end{align}


\end{document}